\documentclass[sn-aps]{sn-jnl}

\jyear{2021}%

\theoremstyle{thmstyleone}%
\theoremstyle{thmstyletwo}%

\theoremstyle{thmstylethree}%

\begin{document}

\title[Supersolid induced by dislocations]{Supersolid induced by dislocations with superfluid cores (Review article)}


\author*[1,2]{\fnm{D. V.} \sur{Fil}}\email{dmitriifil@gmail.com}
\equalcont{These authors contributed equally to this work.}

\author[3]{\fnm{S. I.} \sur{Shevchenko}}\email{shevchenko@ilt.kharkov.ua}
\equalcont{These authors contributed equally to this work.}


\affil*[1]{ \orgname{Institute for Single Crystals of National Academy of Sciences of Ukraine}, \orgaddress{\street{60 Nauky Avenue}, \city{Kharkiv}, \postcode{61072},  \country{Ukraine}}}

\affil[2]{ \orgname{V.N. Karazin Kharkiv National University}, \orgaddress{\street{ 4 Svobody Square}, \city{Kharkiv}, \postcode{61022},  \country{Ukraine}}}

\affil[3]{ \orgname{B.~Verkin Institute for Low Temperature Physics and Engineering, National Academy of Sciences of Ukraine}, \orgaddress{\street{47 Nauky Avenue}, \city{Kharkiv}, \postcode{61103},  \country{Ukraine}}}


\abstract{Dislocation model of the supersolid  state of $^4$He was proposed in 1987 by one of the
authors of the review. The model obtained strong support by numerous experimental and
theoretical investigations from 2007 to date. In these investigations the validity of the
idea put forward in 1987 was confirmed, and new conceptions of superclimb of dislocations
and of the giant isochoric compressibility or the syringe effect were proposed. In this
paper we review main achievements of theoretical and experimental studies of
dislocation-induced supersolid and  present current understanding of this phenomenon.}




\maketitle


\section{INTRODUCTION}

It is known that  quantum liquids become superfluid at low temperature. In 1969 Andreev
and Lifshitz considered properties of quantum crystals (crystals with a large amplitude
of the zero-point oscillations like solid helium) and put  forward the idea on that such
crystals can also exhibit superfluidy \cite{1}. It was shown that at zero temperature
zero-point defectons (vacancies) may exist, and at sufficiently low temperature a gas of
such vacancies becomes a superfluid.

 In 1970 Chester \cite{2} analyzed the conditions under which the many-body states of a system of
interacting bosons may exhibit simultaneously the crystalline order and the Bose-Einstein
condensation. It would seem that the existence of such states contradicts the proof by
Onsager and Penrose \cite{3}  that a state with crystalline order cannot have a
Bose-Einstein condensate in the zero-momentum state. Chester concluded that the proof
\cite{3} is based on the assumption that each particle is localized on a lattice site and
each site is occupied. In the presence of vacancies the latter restrictions is removed,
then the original proof fails.

 The word
"supersolid" was proposed  in 1970 by Matsuda and Tsuneto \cite{4}, who obtained the
criterion of the coexistence of the diagonal and off-diagonal long-range order in the
lattice model of hard-core bosons. Using this criterion,  Matsuda and Tsuneto concluded
that the bulk solid $^4$He is unlikely to become a supersolid, whereas an adsorbed $^4$He
film may become a supersolid. Independently, the term "supersolid" was put forward by
Mullin \cite{5} who considered a cell model of a Bose system. He showed that the model
has three possible phases: normal crystal, Bose-condensed liquid, and a supersolid phase,
having both crystalline order and a Bose condensation. It was proposed that the
supersolid phase might be detected in $^4$He near the liquid-solid transition line.

In 1969 the phenomenon of supersolidity was  discussed in short by Thouless in the paper
devoted to the theory of dense Bose fluids \cite{6}.  Thouless pointed out that a lattice
gas model of dense Bose fluids considered by Matsubara and Matsuda \cite{7} represents,
in fact, a solid with vacancies, and there could be a finite concentration of vacancies
in equilibrium at zero temperature. These vacancies would be in the lowest Bloch state
with a finite probability, and while the system would not be a fluid, it would exhibit
some superfluid features. In particular, if the solid were formed in a ring, the
circulation of vacancies round the ring would be quantized.

The Andreev and Lifshitz suggestion on possible superfluidity of vacancies \cite{1}
stimulated  attempts  to observe superflow in solid $^4$He. The idea of the experiment by
Andreev \textit{et al} \cite{8} was to measure the velocity of a Co-Pt ball frozen in a
helium crystal under the action of artificial gravity created by a superconducting
magnet. However, no measurable motion of the ball was observed. The resulting upper
estimate for the vacancy concentration was less than 0.1\%. In a similar experiment by
Suzuki with a frozen ball \cite{9}, an attempt was made to determine the contribution of
vacancies to the plastic flow of solid helium. It was concluded that the plastic flow can
be described exclusively in the framework of dislocation motion, while the
  vacancy mechanism of such flow is incompatible with the experiment \cite{9}. The study
  of plastic deformation under extremely low loads was carried out by Tsymbalenko \cite{10}.
   The data obtained
  in Ref. \cite{10} also allowed one to conclude that the plastic deformation of crystalline helium is
  of the   dislocation nature.

Plastic properties of solid parahydrogen were investigated by Alekseeva and Krupksii in
\cite{11}. It was found that the plasticity of parahydrogen crystals increases
significantly at the temperature $T=1.8 - 2$ K. In this temperature range,
 the samples were not destroyed when they were deformed up to 50\%.
 After pulling, the samples exhibited damped transverse oscillations.
 The free-standing samples flowed downward in the gravitational field.
 It was suggested in Ref. \cite{11} that this behavior was due to the presence of vacancies.
 The effect \cite{11} was described by Shevchenko \cite{12}. The theory \cite{12} is based
 on the assumption  that zero-point vacancies emerge in the surface layer
 of the crystal of parahydrogen.

Supersolidity would manifest itself in the appearance of the nonclassical moment of
inertia in a cell filled with solid helium. It can be registered by measuring of
temperature dependence of the period of a torsional oscillator. The experiment with the
torsional oscillator was performed by Bishop \textit{et al} \cite{13} in 1981, but the
result was negative. It was concluded in Ref. \cite{13} that either the ratio of the
superfluid density to the total density is less than $5\cdot 10^{-6}$, or the superfluid
transition temperature is less than 25 mK, or the critical superfluid velocity is less
than 5 $\mu$m/s.

An important milestone in the study of supersolidity was the Kim and Chan experiment with
the torsional oscillator  \cite{14} in 2004. In Ref. \cite{14}, the torsional oscillator
with a porous Vycor glass inside was filled with solid helium. A sharp drop in the
oscillation period was observed below 200 mK. The result \cite{14} was reproduced in
several laboratories \cite{15, 16,17,18}. The assumed drop in the moment of inertia of
solid helium in the experiment \cite{14} was 1\%. In other experiments, this drop was
from 0.01\% to 20\%. At the same time, the direct measurement by Day and Beamish
\cite{19} of the shear modulus $\mu$ in solid $^4$He showed that the drop of the period
of the torsional oscillator in the experiments with solid helium is caused, at least,
partially, by the increase of $\mu$.

In the experimental configuration \cite{14} a small amount of solid helium was present
inside the torsion road. Another experiment by the  Chan's group \cite{20} was set up in
a way to exclude any effect of changes in the shear modulus of solid helium. In contrast
with the previous result \cite{14}, no measurable drop in the period of the torsional
oscillator was observed in Ref. \cite{20}. It meant that there was no significant bulk
superfluid fraction in the studied samples.

The supersolid state was originally associated with the presence of zero-point vacancies
in the bulk of the crystal. The concentration of such vacancies should remain finite up
to the temperature $T=0$. But it contradicts to the results of calculations of the
formation energy of the vacancies in solid helium by Boninsegni \textit{et al} \cite{21}.
This formation energy is rather large ($\approx 13$ K) that excludes the existence of
zero-point vacancies in crystals without extended defects.

Apart from vacancies the crystal may contain other kinds of defects, in particular,
different types of dislocations.  The density of dislocations depends on the conditions
of growth of the crystals and can be varies in a wide range. Dislocations are present in
crystals at any temperature, down to $T=0$. In 1987 Shevchenko \cite{22} predicted the
possibility of transition of dislocation cores into a liquid state and superfluidity
along the dislocations at low temperature. The manifestation of the dislocation
supersolid essentially depends on whether dislocations run through the whole crystal
without mutual intersections, or they form a spatial network. In the first case one deals
with a one-dimensional superfluidity along the dislocation line, and in the second case
the superfluidity would exhibit a two-dimensional or three-dimensional character.

In 2007 the superfluid nature of screw dislocations in $^4$He crystals was proven
numerically  by Boninsegni \textit{et al} \cite{23}. The same result for edge
dislocations was obtained in 2009 by Soyler \textit{et al} \cite{24}.

In Ref. \cite{22},  experiments to observe superfluidity along dislocations were
proposed. The contribution of superfluid dislocations to the change in the moment of
inertia is negligible, but the effects due to such superfluidity can be observed in
experiments on the flow of helium atoms through a solid sample. In particular, in
\cite{22} it was suggested the experiment with a $^4$He crystal in contact with He-II. In
a situation where crystalline He separates two regions with liquid He-II, one can observe
a superfluid flow through the crystal. The crystal of $^4$He should be thin enough to
provide that most of the dislocation lines run through the sample. It was also predicted
in Ref. \cite{22} that dislocation superfluidity would be accompanied by superheat
conductivity along dislocation cores, and by anomalous plasticity - superplasticity
similar to that observed in parahydrogen crystals \cite{11}.

 The first attempt to
detect superfluidity along dislocations in $^4$He crystals was done by Bonfair
\textit{et al} \cite{25} in 1989. The experiment \cite{25} was performed to observe the
mass transfer through a crystal enclosed between regions filled with superfluid liquid
helium. The system was kept at the melting curve of the $^4$He phase diagram. When the
solid had grown up in the cell of special shape, the cell was divided in two parts, inner
and outer. More helium can only be admitted in the outer part through a filling
capillary. Further growth in the inner part needed some mass flow through the solid. The
levels of the solid in the inner and outer part are recorded to observe the flow. No mass
transfer was registered down to the temperature $T=4$ mK. It was concluded that above
this temperature the supersolid state does not occur. More precisely, it was established
that the superfluid density in this temperature range does not exceed $10^{-8}$ of the
total density.

The increase of interest to the problem of supersolid since 2004 stimulated further
experiments on the mass flow through $^4$He crystals.   The first successful experiment
in this direction was done by Ray and Hallock \cite{26}. They  used a "sandwich"
consisted of solid helium held between two Vycor plugs connected to two tanks with
superfluid helium. Helium in the pores of Vycor, or other small pore geometries, freezes
at much higher pressures than does bulk helium. It allowed to study the superflow through
the hcp solid $^4$He at pressures off the melting curve under chemical potential gradient
applied across the solid. Atoms were injected into one tank of liquid helium. The
pressure in this tank increased, and the pressure change in another tank of liquid
helium, connected to the first tank through a $^4$He crystal and two Vycor channels, was
measured.  An increase in pressure in the other tank was accounting for the flow of atoms
through the crystal. The value of the flux was estimated from the rate of pressure
change.

There are number of review papers devoted to the phenomenon of supersolidity
\cite{27,28,29,30,31,32,33,34}. In Ref. \cite{27},  experiments on searches of
manifestations of supersolidity as to 1990 were reviewed. The review \cite{28} describes
the progress
 in theoretical and experimental study achieved during the period from 2004 to 2013 and
 stimulated by the experiment \cite{14}. A review of the experiments on mass flow through
solid $^4$He was done in Ref. \cite{29}. The papers \cite{30} and \cite{31} are the
excellent colloquium-style reviews on supersolid. Refs. \cite{32} and \cite{33} are
written for a very broad audience of scientists and describe the main achievements and
problems in understanding of the phenomenon of supersolidity. A short historical review
\cite{34} presents overall portrait and basic ideas related to the problem of supersolid.

 This review is devoted to the problem of dislocation-induced supersolid and phenomena
 related to the presence of superfluid dislocations in quantum crystals.

\section{Superfluid transition in the dislocation network}

\subsection{The idea on the dislocation supersolid}

The idea of Ref. \cite{22} was based on simple physical considerations. It is known that
at the temperature lower than $T_\lambda = 2.17$ K, $^4$He crystallizes at the pressure
about $25$ bar. In real $^4$He crystals, in which dislocations are always present, the
existence of the critical pressure may reveal itself as follows. Let one deals with an
edge dislocation for which an extra half-plane is inserted into the upper part of the
crystal. Then, the upper part of the dislocation core will be in a more compressed state,
and the lower part of the core will be in a less compressed state. At the external
pressure close to the critical one, the pressure in the upper part of the core can be
larger than the critical one, while, in the lower part of the core, it can be smaller
than the critical one. For this reason, the lower part of the dislocation core can be in
a liquid state.

The dislocation is a one-dimensional (1D) object. Therefore, it is instructive to discuss
first the specifics connected with the 1D nature of the superflow along dislocations.

When studying the problem of superfluidity in low-dimensional systems, one is faced with
the question on what a superfluid system is. At first sight, it seems natural to identify
a superfluid system with a system where persistent flow can occur. However, in reality,
such systems do not exist. For any flow with a finite velocity $v_s$ there exists a
finite relaxation time $\tau$. In the general case, this time is a function of the
velocity $v_s$. But two different situations are possible. In some systems the relaxation
time $\tau$ remains finite at $v_s\to 0$. It is natural to consider such systems as
normal ones. In other systems the time $\tau$ is a non-analytic function of $v_s$, and it
approaches infinity at $v_s=0$. The latter systems should be classified as superfluids,
and they should be characterized by a certain order parameter. It is known that
singularities of the order parameters are the source of the relaxation of the superflow.
In three-dimensional (3D) systems the structure with the singularity of the order
parameter is a vortex ring. In two-dimensional (2D) systems, it is a bound pair of
vortices of opposite vorticities (such a pair can be also be considered as a
cross-section of a vortex ring). In 1D systems the singularities of the order parameter
are the phase slip centers (PSCs). The PSC is an instantoneous object localized in the
space and in the time, in which the superfluid density turns to zero.

In 2D system
 the relaxation is connected with  unbinding of vortexes. The binding energy
of the vortex-antovortex pair increases with decreasing the velocity $v_s$ ($E_b\propto
\ln [\hbar/(m\xi v_s)]$, where $\xi$ is the coherence length, and $m$ is the mass of the
Bose particle). Therefore, in 2D systems the relaxation time $\tau \propto \exp(E_b/T)$
goes to infinity at $v_s \to 0$, and the 2D Bose gas becomes superfluid at low
temperature. Similarly, one can show that  $\tau \to \infty$ at $v_s\to 0$ in 3D systems.
In contrast, in the 1D system the energy of the PSC is finite and independent of $v_s$ in
the limit $v_s\to 0$. It follows from the fact that it is sufficient to destroy the
superfluid density at the length of the order of the coherence length to form a PSC.
 Therefore, 1D superfluidity is not possible in a strict sense.
Note that the behavior of the relaxation time at $v_s\to 0$ is in accordance with the
fact that the transition to the superfluid state occurs as a phase transition in 2D and
3D, whereas, according to general theorems, the phase transitions cannot occur at $T\ne
0$ in the 1D case.

A dislocation network is a 3D object. At the same time it possesses some properties of 1D
systems. It results in a dependence of the superfluid transition temperature on the
average length of the segment  of the network.

\subsection{Thermodynamic description of a 1D nonideal  Bose gas}

In this and the next subsection basing on the analysis of a 1D Bose system  we estimate
the superfluid transition temperature in the network. We will follow the paper \cite{35}.

The properties of a 1D system of spinless bosons with delta-function interparticle
interaction is described by the Hamiltonian
\begin{eqnarray}\label{1}
    H=\int_0^{L} dx \Bigg[-\frac{\hbar^2}{2 m}\psi^+(x)\frac{d^2 \psi(x)}{dx^2}-\mu
    \psi^+(x) \psi(x)\cr
    +\frac{\gamma}{2}\psi^+(x)\psi^+(x)\psi(x)\psi(x)\Bigg],
\end{eqnarray}
where  $\psi^+$ and $\psi$ are the Bose creation and annihilation operators,  $\mu$ is
the chemical potential,  $\gamma$ is the interaction constant, and $L$ is the length of
the system. A number of exact results was obtained for the model (\ref{1}) with periodic
boundary conditions. Lieb and Liniger \cite{36} have calculated the ground state energy
and the spectrum of elementary excitations for this model. Yang and Yang \cite{37} have
shown that thermodynamic functions of this model are analytical at any finite
temperature, which indicates the absence of a phase transition.

Finite systems may demonstrate a number of regimes of quantum degeneracy. In particular,
three regimes of quantum degeneracy in a trapped 1D gas were identified by Petrov,
Shlyapnikov, and Walraven \cite{38}: the true Bose-Einstein condensate, a
quasicondensate,  and a Tonks gas. As low density the 1D gas of interacting Bose
particles acquires Fermi properties, and it can be described as a gas of impenetrable
bosons (the Tonks gas), whereas at high density the mean-field  theory of superfluidity
can be used.

The partition function of the model,  $Z=\mathrm{Tr}\left[\exp\left(-\beta
H\right)\right]$, where $\beta=1/T$, can be written in the functional integral form
\begin{equation}\label{2}
    Z=\int D\psi^*(x,\tau)D\psi(x,\tau)e^{S(\beta)},
\end{equation}
where the action $S$ is given by the equation
\begin{equation}\label{3}
    S(\beta)=\int_0^\beta d\tau\left[\int_0^L d x
    \psi^*(x,\tau)\frac{\partial \psi(x,\tau)}{\partial \tau}-H(\tau)\right],
\end{equation}
and $H(\tau)$ is obtained from the Hamiltonian (\ref{1}) by the substitution $\psi(x)\to
\psi(x,\tau)$ and $\psi^+(x)\to \psi^*(x,\tau)$. Eqs. (\ref{2}) and (\ref{3}) are exact.

The mean-field approximation corresponds to the reduction of the functional integral
(\ref{2}) to the Gauss integral. The order of magnitude of the action (\ref{3}) is $\beta
\gamma n^2 \xi$, where $n$ is the average 1D density of the particles, $\xi=\hbar/\sqrt{2
m\gamma n}$ is the coherence length. At low temperature the quantity $\beta \gamma n^2
\xi$ is much larger than unity, the exponent in Eq. (\ref{2}) is large, and the main
contribution to the partitions function comes from the functions $\psi^*(x,\tau)$ and
$\psi(x,\tau)$ at which the action $S(\beta)$ reaches the extremum. The extremum
condition is given by the equation

\begin{eqnarray}\label{4}
 \frac{\partial \tilde{\psi}(x,\tau)}{d \tilde{\tau}}+\frac{\partial^2 \tilde{\psi}(x,\tau)}
 {d \tilde{x}^2}
 + \tilde{\psi}(x,\tau)-\lvert\tilde{\psi}(x,\tau)\rvert^2 \tilde{\psi}(x,\tau)=0,
\end{eqnarray}
where the dimensionless variables $\tilde{\psi}=\psi/\sqrt{n}$, $\tilde{\tau}=\gamma n
\tau$, and $\tilde{x}=x/\xi$ are used.  In obtaining Eq. (\ref{4}) it was taken into
account that the chemical potential is equal to $\mu=\gamma n$.

The minimum of $S(\beta)$ is reached at
$\tilde{\psi}=\tilde{\psi}_0=1$. To obtain the contribution of small
deviations of $\tilde{\psi}$ from $\tilde{\psi_0}$ to the partition
functions, it is convenient to write the function $\psi(x,\tau)$ in
the form
$\psi(x,\tau)=e^{i\varphi(x,\tau)}\sqrt{n+\delta\rho(x,\tau)}$ and
take into account the quadratic in $\delta\rho(x,\tau)$ and
$\varphi(x,\tau)$ corrections to the extremum $S(\beta)$. The
Fourier-components of $\delta\rho(x,\tau)$ and $\varphi(x,\tau)$ are
presented in the form
\begin{eqnarray} \label{5}
 \delta\rho(k,\omega_N)&=&  \sqrt{\frac{\epsilon(k) n}{E(k) L}}
 \left[\eta(k,\omega_N)+\eta^*(-k,-\omega_N)\right],\cr
  \varphi(k,\omega_N) &=& \frac{1}{2i}\sqrt{\frac{E(k)}{\epsilon(k) n L}}
   \left[\eta(k,\omega_N)-\eta^*(-k,-\omega_N)\right],
\end{eqnarray}
where $\epsilon(k)=\hbar^2 k^2/2m$ is the spectrum of noninteracting bosons,
$E(k)=\sqrt{\epsilon(k)[\epsilon(k)+2\gamma n]}$ is the Bogoliubov spectrum, and
$\omega_N=2\pi N T$ ($N$ is integer) are the Matsubara frequencies. The action is
diagonal in the variables $\eta(k,\omega_N)$:
\begin{equation}\label{6}
    S(\beta)=\frac{1}{2}\gamma n^2 L \beta+\beta\sum_{k,\omega_N}\left[i\omega_N+E(k)\right]
    \eta^*(k,\omega_N) \eta(k,\omega_N).
\end{equation}
The partition function,
\begin{equation}\label{7}
    Z=\prod_{k,\omega_N}\int d \eta^*(k,\omega_N) d \eta(k,\omega_N) e^{S(\beta)}
\end{equation}
coincides with the partition function of a gas of noninteracting particles with the
spectrum $E(k)$. The free energy is equal to
\begin{equation}\label{8}
    F=-T \ln Z =-\frac{1}{2}\gamma n^2 L-T\sum_k \ln\left[1-e^{-\beta E(k)}\right].
\end{equation}
The free energy (\ref{8}) is an analytical function of temperature, and, therefore, no
phase transitions occur in the system as the temperature varies. However, within this
approximations, i.e., when only small fluctuations of $\psi(x,\tau)$ near the extreme
value of $\psi$ are allowed, the system behaves as a superfluid one. Indeed, when the
walls move with the velocity $v$, one needs to add  the term $-\int_0^L d x j v$ to the
Hamiltonian, where
\begin{equation}\label{9}
    j=\frac{i\hbar}{2 } \left(\psi\frac{d \psi^*}{d x}-c.c.\right)
\end{equation}
is the mass flow.

 The thermodynamically average mass flow  is given
by the expression
\begin{equation}\label{10}
\langle j \rangle=\frac{1}{Z}\int D \psi^* D \psi \frac{i\hbar}{2 } \left(\psi\frac{d
\psi^*}{d x}-c.c.\right) e^{S_v(\beta)},
\end{equation}
where $S_v(\beta)$ is the action that takes into account the term $-\int_0^L d x j v$.
The calculation of the average (\ref{10}) yields
\begin{equation} \label{11}
\langle j \rangle =- \frac{v}{2\pi} \int_{-\infty}^{\infty} d k \hbar^2 k^2 \frac{d
N_B(E)}{d E}\vert_{E=E(k)},
\end{equation}
where $N_B(E)=(e^{\beta E}-1)^{-1}$ is the Bose distribution function.  Using the
relation $\langle j \rangle=m \rho_n v$ and Eq. (\ref{11}) one can calculate the normal
density $\rho_n$. At $T\ll \gamma n$ Eq. (\ref{11}) yields
\begin{equation}\label{12}
    \rho_n=\frac{\pi T^2}{3m\hbar c^3},
\end{equation}
where $c=\sqrt{\gamma n/m}$  is the velocity of the sound mode. In fact, Eq. (\ref{12})
is the Landau formula for the normal density of a 1D superfluid. According to Landau, at
sufficiently low temperature the normal density $\rho_n$ is smaller than the total
density $\rho$, and the superfluid density $\rho_s=\rho-\rho_n$ is nonzero. But it does
not contradict the statement on the absence of superfluidity in 1D systems at $L\to
\infty$.  Due to nonzero probability of the emergence of PSC, the superfluid density
turning to zero at some points. In a system of the infinite length, PSC are always
present that leads to damping of the flow.

\subsection{Temperature of the superfluid transition in the network}

Let us now return to the dislocation model of the supersolid state of $^4$He crystal.
Bosons in a 3D and even in a 2D network  undergo the phase transition to the superfluid
state as the temperature decreases. The  temperature of phase transition does not depend
on the characteristic size of the system $L$, but it may depend on the average length of
the segment of the network $l$. The relation between the  temperature of phase transition
and the length $l$ can be found with the use of the 1D correlation function
\begin{equation}\label{13}
    g(x)=\langle \psi^*(x) \psi(0)\rangle.
\end{equation}
Neglecting the density fluctuations and using the approximation (\ref{6}),  one obtains
\begin{equation}\label{14}
   g(x)=\frac{1}{Z}\int D \varphi n e^{-i[\varphi(x)-\varphi(0)]+S(\beta)}\approx
   ne^{-\frac{\lvert x\rvert}{\lambda(T)}},
\end{equation}
where
\begin{equation}\label{15}
    \lambda(T)=\frac{2\hbar^2 \rho_s(T) }{m T}
\end{equation}
is the decay length of phase correlations. The phase coherence in the network is
established at $\lambda(T)\gtrsim l$. The temperature of the superfluid transition in the
network is obtained from the equation $\lambda(T_c)\approx l$, or
\begin{equation}\label{16}
    T_c \approx \frac{\hbar^2 \rho_s(T_c)}{m l}.
\end{equation}

\begin{figure}
\begin{center}
\includegraphics[width=8cm]{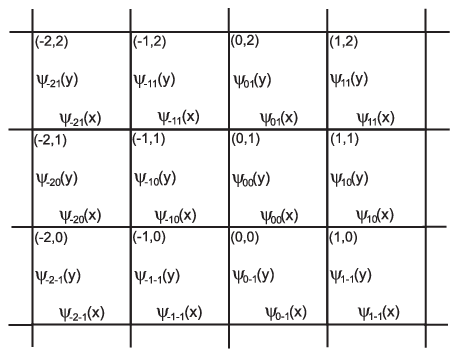}
\end{center}
\caption{Node and order parameter functions in the periodic 2D network.}\label{f1}
\end{figure}

More careful analysis can be done for the periodic 2D network which nodes form the
quadratic lattice with the period $l$. The nodes can be labeled as $(s,p)$, where the
integer $s$ and $p$ are the $x$ and $y$ coordinates of the nodes in units of $l$ (see
Fig. \ref{f1}). The order parameter functions $\psi(x)$ at the segments which connect the
$(s,p)$ and $(s+1,p)$ nodes are labelled as $\psi_{s p}(x)$, and the functions $\psi(y)$
at the segments which connect the $(s,p)$ and $(s,p+1)$ nodes are labelled as $\psi_{s
p}(y)$.  These functions are sought in the form
\begin{eqnarray}\label{17}
    \psi_{s p}(x)=C^{(x)}_{s p} \exp\left(i k^{(x)}_{s p} x\right), \cr \psi_{s p}(y)
    =C^{(y)}_{s p} \exp\left(i k^{(y)}_{s p} y\right),
\end{eqnarray}
where $k^{(x)}_{s p}=(\varphi_{s+1,p}-\varphi_{s,p})/l$, $k^{(y)}_{s
p}=(\varphi_{s,p+1}-\varphi_{s,p})/l$, and $\varphi_{s,p}$ is the phase of the order
parameter at the $(s,p)$ node. Substituting Eqs. (\ref{17}) into Eq. (\ref{4}) one finds
that $C_{s p}^{(x,y)}=\sqrt{n\left[1-\xi^2 \left(k^{(x,y)}_{s p}\right)^2\right]}$.

The superfluid phase transition in 2D is connected with unbinding of vortex pairs.
Vortexes may emerge as fluctuations. Eqs. (\ref{17}) describes the state with quantum
vortexes as well. The essential difference between quantum vortexes in a continuous
medium and in the network is that in the latter case the vortexes have no normal core:
they are pure circular supercurrents.
 Such currents decay by the law $1/r$, where $r$
is the distance from the vortex center, and, therefore,  the inequalities $k^{(x,y)}_{s
p}\ll \xi^{-1}$ are satisfied. Then, one can use the approximation $C_{s
p}^{(x,y)}\approx \sqrt{n}$.

The calculation of the partition function of a segment yields
\begin{equation}\label{18}
    Z^{(x,y)}_{s p}=Z_0\exp\left[-\beta m \rho_s(T) l \frac{v_s^2}{2}\right],
\end{equation}
where $v_s=\hbar k^{(x,y)}_{s p}/m$ is the superfluid velocity, $\rho_s(T)$ is the 1D
superfluid density at the segment, and $Z_0$ is the partition function (\ref{7}). Then,
the partition function of the network is presented in the form
\begin{equation}\label{19}
    Z=Z_0 \int D \varphi_i \exp\left[-\beta \frac{J}{2}\sum_{\langle  ij\rangle}
    (\varphi_i-\varphi_j)^2\right],
\end{equation}
where $J=\hbar^2 \rho_s(T)/m l$, the index $i$ ($j$) is the node label, and the summation
is over nearest neighbor pairs of nodes. Up to the regular factor $Z_0$, and at small
difference $\varphi_i-\varphi_j$, the function (\ref{19}) coincides with the partition
function of the 2D $xy$-model. As is known, the latter demonstrates the
Berezinskii-Kosterlits-Thouless (BKT) phase transition at the temperature $T_c=\pi J/2$.
Thus, the superfluid transition temperature in the 2D network is given by equation
\begin{equation}\label{20}
    T_c=\frac{\pi}{2}\frac{\hbar^2 \rho_s(T_c)}{m l}.
\end{equation}
One can see that up the the factor $\pi/2$ this equation coincides with Eq. (\ref{16}).

The same procedure for the 3D periodic network gives the partition function of the 3D
$xy$-model. The critical temperature for the bcc $xy$-model, obtained numerically
\cite{39}, is equal to $T_c\approx 2.2 J$. Thus, the superfluid transition temperature in
the cubic network satisfies the equation
\begin{equation}\label{21}
    T_c\approx 2.2\frac{\hbar^2 \rho_s(T_c)}{m l}.
\end{equation}
Up to the numerical factor it coincides with Eq. (\ref{20}).

\subsection{Bose-Einstein condensation of an ideal Bose gas in a decorated lattice}

The periodic 3D dislocation network can be modelled by a decorated lattice. In Ref.
\cite{40}  by Fil and Shevchenko, this model was considered to evaluate the temperature of the Bose-Einstein
condensation of an ideal gas of Bose particles in such a network. It was found that  in a
special case the model yields the same as above  dependence of $T_c$ on $l$ ($\propto
1/l$). In this subsection we present the results of Ref. \cite{40}.

\begin{figure}
\begin{center}
\includegraphics[width=4cm]{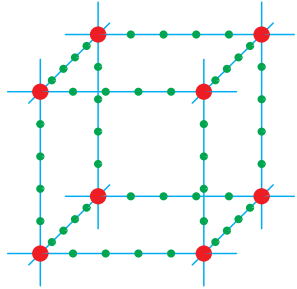}
\end{center}
\caption{Elementary cell of the decorated lattice with $q=5$. Large red circles represent
node sites,  small green circles represent decorating sites.}\label{f2}
\end{figure}

The decorated lattice considered in Ref. \cite{40} has two types of lattice sites, the
node sites which form a cubic lattice with the period $l$, and the decorating sites which
form chains connecting the node sites (Fig. \ref{f2}). The period of the chain is $a$.
The number of sites in the elementary cell is equal the $3q-2$, where $q=l/a$. The gas of
noninteracting bosons  was described in the tight-binding approximation. Only nearest
neighbor hopping was taken into account. The amplitude of hopping between two decorating
sites ($t$) and the amplitude of hopping between the node site and the decorating site
($t_1$) were assumed to differ from each other. The same one-site energies were taken for
all sites. The number of bosons was supposed to be much smaller than the overall numbers
of sites, but much larger than the number of the node sites.

The Hamiltonian of the model has the form
\begin{equation}\label{22}
    H=\sum_{\mathbf{k},\eta_1,\eta_2} G_{\eta_1,\eta_2}(\mathbf{k}) b^+_{\eta_1,\mathbf{k}}
    b_{\eta_2,\mathbf{k}},
\end{equation}
where $b^+_{\eta,\mathbf{k}}$ and $b_{\eta,\mathbf{k}}$ are Bose creation and
annihilation operators, the index $\eta$ enumerates $3q-2$ sites in the unit cell,
$\mathbf{k}$ is the wave vector, and $\mathbf{G}(\mathbf{k})$ is the $(3q-2)\times(3q-2)$
matrix of the form
\begin{equation}\label{23}
    \mathbf{G}({\bf k})=-t\left(
\begin{array}{cccc}
  0 & \mathbf{T}_x & \mathbf{T}_y  & \mathbf{T}_z  \\
  \mathbf{T}_x^+  & \mathbf{D}_{q-1} & \mathbf{0} & \mathbf{0} \\
   \mathbf{T}_y^+ & \mathbf{0}  & \mathbf{D}_{q-1}  & \mathbf{0}  \\
   \mathbf{T}_z^+  & \mathbf{0}  & \mathbf{0}  & \mathbf{D}_{q-1} \\
\end{array}
\right).
\end{equation}
In Eq. (\ref{23}),
\begin{equation}\label{24}
\mathbf{D}_{q-1}=\left(%
\begin{array}{ccccc}
  0 & 1 & \ldots & 0 & 0 \\
  1 & 0 & \ldots & 0 & 0 \\
  \ldots & \ldots & \ldots & \ldots & \ldots \\
  0 & 0 & \ldots & 0 & 1 \\
  0 & 0 & \ldots & 1 & 0 \\
\end{array}%
\right)
\end{equation}
is the $(q-1)\times(q-1)$ matrix, and
$$ \mathbf{T}_\alpha=\left(%
\begin{array}{ccccc}
  \tau & 0 & \dots & 0 & \tau e^{-i k_\alpha l} \\
\end{array}%
\right)$$  are the $1\times(q-1)$ matrixes ($\alpha=x,y,z$).  Here we introduce the
parameter $\tau=t_1/t$. The matrix $\mathbf{D}_{q-1}$ describes the tunneling between the
decorating sites, and the matrixes $ \mathbf{T}_\alpha$ described the tunneling
 between  the node  and decorating sites.

The dispersion equation for the spectrum of bosons is obtained from the equation $\det[
\mathbf{G}({\bf k})-E\mathbf {I}_{3q-2}]=0$, where  $\mathbf {I}_n$ is the identity
matrix of the dimension $n$. The dispersion equation reads
\begin {eqnarray} \label{25}
    [\Delta_ {q-1} (\tilde{\epsilon})] ^2 [\tilde{\epsilon} \Delta_ {q-1} (\tilde{\epsilon})-6
    \tau^2 \Delta _ {q-2} (\tilde{\epsilon}) \cr - 2(-1)^q  \tau^2 \sum_{\alpha=x,y,z}
\cos (k_\alpha l)] =0,
\end {eqnarray}
where $\tilde{\epsilon}=E/t$ is the energy normalized to $t$, and $\Delta _ {q}
(\tilde{\epsilon}) = \det (\tilde{\epsilon} \mathbf {I}_q + \mathbf {D} _ {q})$.

The spectrum contains $q-1$ degenerate levels which energies are obtained from the
equation
\begin {equation} \label {26}
    \Delta_ {q-1} (\tilde{\epsilon})=0.
\end {equation}

 One finds from the explicit form of $\mathbf {D} _ {q}$
that the function $\Delta_ {q} (\tilde{\epsilon})$ satisfies the recurrence relation
\begin {equation} \label{27}
\Delta_q (\tilde {\epsilon}) = \tilde {\epsilon} \Delta _ {q-1} (\tilde {\epsilon}) -
\Delta _ {q-2} (\tilde {\epsilon})
\end {equation}
with $ \Delta_1 (\tilde {\epsilon}) = \tilde {\epsilon} $ and $ \Delta_2 (\tilde
{\epsilon}) = \tilde {\epsilon} ^2-1$. Using the relation (\ref{27}) and applying the
method of the mathematical induction one can prove  that
\begin {equation} \label {28}
    \Delta_q (2 \cos \gamma) = \frac {\sin [(q+1) \gamma]} {\sin\gamma}.
\end {equation}
From Eqs. (\ref{26}) and (\ref{28}) one  finds the energies of degenerate levels:
\begin {equation} \label {29}
    E_s =-2 t \cos\frac {\pi s} {q},
\end {equation}
where $s=1,2, \ldots, q-1$. The degree of degeneracy of each level is $2N_i$, where $N_i$
is the number of unit cells in the system.

Beside the levels (\ref{29}), there are
 $q$ bands of finite widths. The dispersion law for these bands
 can be found from the equation
\begin {eqnarray} \label{30}
    \tilde{\varepsilon} \Delta_ {q-1} (\tilde{\varepsilon})-6
    \tau^2 \Delta _ {q-2} (\tilde{\varepsilon}) \cr - 2(-1)^q  \tau^2 \sum_{\alpha=x,y,z}
\cos (k_\alpha l) =0.
\end {eqnarray}

\begin{figure}
\begin{center}
\includegraphics[width=6cm]{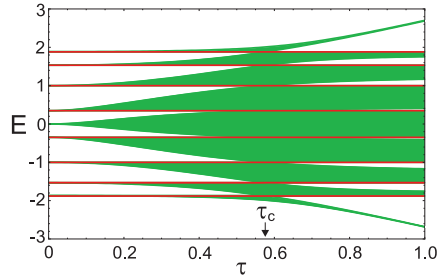}
\end{center}
\caption{Band structure of the decorated lattice model at $q=9$ versus the parameter
$\tau$. The energy $E$ is in units of $t$. The degenerate levels are shown red, and the
finite width bands are shown green.}\label{f3}
\end{figure}

 An example of the band structure at $q=9$  versus the parameter $\tau$ is
shown in Fig. \ref{f3}. One can see that at the special value of $\tau =\tau_c=1/\sqrt
{3}$ the widths of all band reach the maximum, and there is no gaps between the bands.

At $\tau=\tau_c$ Eq. (\ref{30}) reduces to
\begin {eqnarray} \label{31}
   2 \Delta_ {q} (\tilde{\varepsilon})- \tilde{\varepsilon} \Delta_ {q-1} (\tilde{\varepsilon})
    =\frac{2}{3}(-1)^q   \sum_{\alpha=x,y,z}
\cos (k_\alpha l).
\end {eqnarray}
Substituting $\tilde {\epsilon}=2\cos(\pi+\gamma')$ into Eq. (\ref{31}) and taking into
account Eq. (\ref{28}) one finds the equation
\begin{equation}\label{32}
   \cos (q\gamma')=\frac {1} {3} \sum_{\alpha=x,y,z} \cos (k_\alpha l).
\end{equation}
Eq. (\ref{32}) is satisfied with $q$ different values of $\gamma'$ which determine the
spectrum
\begin {eqnarray} \label{33}
    E_ {j} ({\bf k}) =-2t\cos
    \Bigg(\frac {2\pi} {q} \left [\frac {j} {2} \right] \cr -
(-1) ^j\frac {1} {q} \arccos\frac {\sum_{\alpha=x,y,z} \cos
(k_\alpha l)} {3} \Bigg)
\end {eqnarray}
($j=1,2, \ldots q $), where square brackets indicate the integer part. From Eq.
(\ref{33}) one obtains approximate expression for the widths of the bands with $j\ll q $:
\begin {equation} \label {34}
    W_j\approx \frac {t \pi^2} {q^2} (2j-1).
\end {equation}

The temperature of  Bose-Einstein condensation (BEC) is obtained from the equation
\begin {eqnarray} \label {35}
    n_{3D} =\frac{1}{V}\sum _ {j=1} ^ {q} \sum _ {{\bf k}}
    \frac {1} {\exp\left (\frac {E_j (\bf
    k)-\mu_0} {T_c} \right)-1}
    \cr +\frac{2}{l^3}\sum _ {s=1} ^ {q-1} \frac {1} {\exp\left (\frac {E_s-\mu_0} {T_c}
    \right)-1},
\end {eqnarray}
where $\mu_0$ coincides with the minimum of the lowest energy band, and $n_{3D}=3 n/l^2$
is the average 3D density of the bosons ($n$ is the linear density of boson in the
chains).

For the spectrum (\ref{29}), (\ref{34}), at the density $n_{3D}l^3\gg 1$ the temperature
of BEC is evaluated as
$$T_c\approx 0.1 W_1 n_{3D} l^3$$
that yields
\begin {equation} \label {36}
    T_c\approx 3 t  \frac {n a^2} {l}.
\end {equation}
Introducing the effective mass of 1D bosons in the chain, $m_{eff} =\hbar^2/2t a^2$, one
can rewrite Eq. (\ref{36}) as
\begin {equation} \label {37}
 T_c\approx \frac{3}{2} \frac {\hbar^2} {m_{eff}} \frac {n} {l}.
\end {equation}
Up to the numerical factor of order of unity it coincides with the critical temperature
(\ref{21}).

The case $\tau=\tau_c$ is a distinct one since in this case all lattice sites are equally
filled in the ground state. In contract, at $\tau<\tau_c$ a reduced occupation of the
node sites occurs, and at $\tau>\tau_c$ bosons  are located presumably in the node sites.
It results in narrowing of the lowest energy band (see Fig. \ref{f1}) and lowering of
$T_c$. Calculations show that at $\tau<\tau_c$ and $l\gg a/(1-\tau^2/\tau_c^2)$ the
critical temperature $T_c \propto 1/l^2$. At $\tau>\tau_c$ and $l\gg
a/(\tau^2/\tau_c^2-1)$ the critical temperature is exponentially small: $T_c \propto
l\exp(-\alpha l/a)$, where $\alpha$ is numerical factor of order of unity.

From the obtained different dependencies of $T_c$ on the length $l$ one should choose the
result which is stable with respect to  switching on the interaction. Of those three
results  the answer (\ref{36}) satisfies this condition. Thus, we conclude that only the
special case $\tau=\tau_c$ describes adequately the physical situation.

\subsection{Coupling of the superfluid order parameters with strains induced by dislocations}

The Ginzburg-Landau type theory of the dislocation supersolid  was developed by Toner
\cite{41}. The model \cite{41} is based on the following phenomenological Hamiltonian
\begin{equation}
\label{38}
    H=\int d^3 x \left[\frac{t(\mathbf{r})}{2}\lvert\psi \rvert^2
   + \frac{u}{4}\lvert\psi \rvert^4+\frac{c}{2} \lvert\nabla \psi \rvert^2\right],
\end{equation}
where $\psi(\mathbf{r})$ is the superfluid order parameter, $u$ and $c$ are the
phenomenological constants, and the coefficient $t$ is the function of local stains:
\begin{equation}\label{39}
t(\mathbf{r})=t_0(T)+g \sum_{i=x,y,z} u_{ii}(\mathbf{r}),
\end{equation}
Here $t_0(T)$ is the decreasing function of $T$, $u_{ii}$ are the diagonal components of
the strain tensor, and the constant $g$ describes the coupling of elastic strains and the
superfluid order parameter.

The model (\ref{38}) allows one to describe two situations. The first one is a
hypothetical. It corresponds to the case where even a dislocation-free crystal can
transit into the supersolid state. The temperature of such a transition, $T_{c0}$, is
given by the equation $t_0(T_{c0})=0$. Above this temperature superfluidity is possible
along the dislocation cores. The second situation corresponds to the case where the
transition of the whole crystal to the supersolid state is not possible  ($t_0(T)>0$ for
all $T$), but the superfluid order parameter in the dislocation core can be nonzero (we believe that just this situation is realised in solid $^4$He).

For the energy (\ref{38}), the condition $\psi(\mathbf{r})\ne 0$ is equivalent to the
requirement that the Euler-Lagrange equation
\begin{equation}\label{40}
    \nabla^2 \psi=\frac{t(\mathbf{r})}{c}\psi+\frac{u}{c} \psi^3
\end{equation}
has a nontrivial solution.

For a straight edge dislocation running along the $z$ axis with Burgers vector
$\mathbf{b}$ along the $y$ axis,
\begin{equation}\label{41}
  \sum_{i=x,y,z} u_{ii}(\mathbf{r})=\frac{4 \mu}{2\mu+\lambda}\frac{b \cos
  \theta}{r_\perp},
\end{equation}
where $\mu$ and $\lambda$ are the Lame elastic constants, and $r_\perp$ and $\theta$ are
the cylindrical  coordinates in the plane perpendicular to the dislocation line. Then,
the function (\ref{39}) reads
\begin{equation}\label{42}
 t(\mathbf{r})=t_0(T)+g'\frac{\cos \theta}{r_\perp},
\end{equation}
where $g'= 4 g b\mu/(2\mu+\lambda)$. Under neglecting  of the cubic term, Eq. (\ref{40})
coincides in form with the stationary Schroedinger equation for a particle of mass $m$ in
the 2D dipole potential with the strength $p$,
\begin{equation}\label{43}
-\frac{\hbar^2}{2m}\nabla^2 \psi+\frac{p \cos \theta}{r_\perp}\psi=E\psi
\end{equation}
(under substitution $E=-\hbar^2 t_0/(2m c)$ and $p=\hbar^2 g'/(2mc)$). Eq. (\ref{43}) has
the bound state with the energy $E_0=-2x_0 m p^2/\hbar^2$, where the constant $x_0
\approx 0.1$ (see \cite{42}). Accordingly, Eq. (\ref{40}) has nontrivial solutions at
\begin{equation}\label{44}
    t_0< x_0 \frac{(g')^2}{c}.
\end{equation}
This inequality defines the region of parameters at which the supersolid state can be
realized. The meanfield critical temperature is obtained from equation
\begin{equation}\label{45}
    t_0(T_c)=x_0 \frac{(g')^2}{c}.
\end{equation}
If follows from Eq. (\ref{45}) that condensation onto dislocations can happen, even when
the clean system does not order (i.e. at $t_0(0)>0$), but the clean system should be
close to the supersolid transition  ($t_0(0)$ should be smaller than $2x_0
{(g')^2}/{c}$).

Further analysis in Ref. \cite{41} concerned the problem of the superfluid transition in
the network.
 Assuming that the only important variable is the phase $\varphi(x)$ of the
superfluid order parameter, Toner considered the following Hamiltonian for a segment of
the network,
\begin{equation}\label{46}
    H_{1D}(\{\varphi(x)\})=\frac{K}{2}\int_0^l d x \left(\frac{\partial \varphi}{\partial
    x}\right)^2,
\end{equation}
where $K$ is the 1D superfluid stiffness, and $l$ is the length of the segment. The
superfluid stiffness is calculated as
\begin{equation}\label{47}
    K=\frac{\hbar^2}{m}\int d^2 r_\perp \rho_{3s}(\mathbf{r})=\frac{\hbar^2 \rho_{s}}{m},
\end{equation}
where $\rho_{3s}$ is the 3D superfluid density.
 From Eq. (\ref{46}) one obtains the effective Hamiltonian which describes the coupling
of the phases $\varphi_i$ and $\varphi_j$ at the edges of the segment,
\begin{eqnarray}\label{48}
    e^{-\beta H_{eff}(\varphi_i,\varphi_j)}=\sum_{n=-\infty}^{+\infty}\int D \varphi(x)
 e^{-\beta H_{1D}(\{\varphi(x)\})},
\end{eqnarray}
where $\varphi(x)$ satisfies the boundary conditions $\varphi(0)=\varphi_i$ and
$\varphi(l)=\varphi_j+2\pi n$.

Under substitution
\begin{equation}\label{49}
\varphi(x)=\varphi_i+\left(\frac{\varphi_j-\varphi_i+2\pi n}{l}\right)x +\delta
\varphi(x)
\end{equation}
($\delta \varphi(0)=\delta\varphi(l)=0$), the integral in Eq. (\ref{48}) reduces to
\begin{eqnarray}\label{50}
    e^{-\beta H_{eff}(\varphi_i,\varphi_j)}=\sum_{n=-\infty}^{+\infty}
e^{-\frac{\beta K}{2 l}\left(\varphi_j-\varphi_i+2\pi n\right)^2}
  \cr \times  \int D \delta \varphi(x)
 e^{-\frac{\beta K}{2} \int_0^l
 dx \left(\frac{\partial \delta \varphi(x)}{\partial x}\right)^2}.
 \end{eqnarray}
The integral in Eq. (\ref{50}) only adds an irrelevant constant $C$ to $H_{eff}$:
\begin{eqnarray}\label{51}
   H_{eff}(\varphi_i,\varphi_j)=V(\varphi_i,\varphi_j,J)\cr
   =-T \ln\left(\sum_{n=-\infty}^{+\infty}
e^{-\frac{\beta J}{2}\left(\varphi_j-\varphi_i+2\pi n\right)^2}\right)+ C,
\end{eqnarray}
where $J=K/l$. The effective Hamiltonian for the network is given by the sum over bonds
\begin{equation}\label{52}
    H_{eff}= \sum_{\langle i j\rangle}V(\varphi_i,\varphi_j,J).
\end{equation}
This is the Villain model. The inverse critical temperature for the 3D Villain model is
equal to $\beta_c\approx 0.33 J^{-1}$ \cite{43}. It yields the equation for $T_c$:
\begin{equation}\label{53}
    T_c\approx 3 \frac{\hbar^2 \rho_s(T_c)}{m l}.
\end{equation}

If $t_0(0)<0$, the radius of the supersolid area around dislocations diverges at
$T=T_{c0}$, and the critical temperature is equal to $T_c=T_{c0}+\delta T$, where $\delta
T\propto l^{-3/4}$. 

If $t_0(0)>0$, the radius of the supersolid area around the dislocation line remains
finite at $T\to T_c$. In this case the temperature of the superfluid transition is
proportional to $1/l$. Up to the numerical factor of order of unity the obtained critical
temperature coincides with one given by Eq. (\ref{21}).

\subsection{Dislocation network with trapped vorticity}

The role of trapped vortexes in superfluid transition in the dislocation network was
discussed by Nozieres in Ref. \cite{44}, where the following scenario was considered.

The density of $^4$He varies under solidification. If the solidification occurs at fixed
volume, edge dislocations form in large numbers. The dislocations are very mobile. It
results in unusually high plasticity. Under large elastic stresses, a dislocation may
trap vacancies to reduce its energy. In such a picture, the core is softened and although
it is not exactly liquid, it can be in a supersolid state.

Motion of dislocations generates plastic flow, which can lead to
relaxation of applied shear stresses. In an ordinary material, a
dislocation can move only through  drift of kinks along it. Such
movement is slow and plasticity is small. Quantum fluctuations at $T
= 0$ in a 1D system are similar to thermal fluctuations in a 2D
system. In 2D, the surface
 may undergo a roughening transition as a function of $T$ (fluctuations diverge on a large
scale). A 1D system  may undergo similar transition (bending) at $T
= 0$ as a function of line stiffness. Therefore, plastic flow of
$^4$He crystals occurs easily. Even the concept of kinks may lose
its meaning.

The decrease in shear elasticity is very sensitive to $^3$He
impurities. It is energetically preferable for these impurities to
locate in dislocation cores. At low temperatures (tens of mK)
dislocations are fixed by impurities and do not move. At higher
temperature dislocations detach from impurities and the shear
modulus decreases. The distribution of capture centers is random,
and when the applied strain increases, uncoupling from attachment
points does not occur simultaneously throughout the crystal. As a
result, dislocations form a complex network of intersecting loops
filled with superfluid phase.

Nozieres proposed the following simple derivation of the dependence $T_c\propto 1/l$. The
supersolid state is characterized by phase coherence but phase fluctuations in a 1D
system diverge. To see this one can write the kinetic energy in terms of the
Fourier-component of the phase:
\begin{equation}\label{54}
    E=\int_0^l d x \frac{m \rho_s v_s^2}{2}=\frac{\hbar^2 \rho_s }{2 m}l\sum_q q^2
    \lvert\varphi_q\rvert^2.
 \end{equation}
The energy that corresponds a given $q$ satisfies the equipartition law,
\begin{equation}\label{55}
    \frac{T}{2}=\frac{\hbar^2}{2m}\rho_s l q^2\lvert\varphi_q\rvert^2.
\end{equation}
The square of difference of the phase fluctuations at the ends of the system of the
length $l$ is given by the expression
\begin{equation}\label{56}
    \lvert\Delta\varphi\rvert^2=2\sum_{q}\lvert\varphi_q\rvert^2
    \left[1-\cos(q l)\right].
\end{equation}
Substituting $\lvert\varphi_q\rvert^2$ from Eq. (\ref{55}) into Eq. (\ref{56}) one finds
\begin{equation}\label{57}
    \lvert\Delta\varphi\rvert^2=\frac{m T l}{\hbar^2  \rho_s}.
\end{equation}
It follows from  Eq. (\ref{57}) that at some critical temperature the mean square
fluctuation of the phase difference will be equal to $2\pi$. This critical temperature
corresponds to the loss of phase coherence and its value is inversely proportional to
$l$.

However, according to Nozieres, it is possible to introduce another critical temperature.
A tangle of dislocation loops will necessarily contain a lot of trapped vortexes. Each
loop may have its own quantum circulation number which depends on the past history. If
the loop is stationary and there are no weak bonds in it, such a trapped vortex does not
greatly affect the superfluid flow in the sample: the latter is just superimposed to the
vortex pattern. This is what happens at low temperatures when dislocations are fixed on
$^3$He impurities. At a higher temperature, the dislocations begin to move. At some
critical temperature, when the displacements are compared to the typical length of the
segment $l$, the connectivity of the tangle changes: any trapped vorticity is
redistributed in another way. This redistribution leads to phase noise, which destroys
the non-diagonal long-range order. In such a picture, the critical temperature is
associated not with local superfluidity, but with the onset of plasticity, which blurs
the underlying vorticity.

\section{Dynamical model of quasisuperflow in the dislocation network}
\label{s3}

 In Refs. \cite{35,45},  unusual dynamical properties of a network of
superfluid dislocations was predicted. It was shown that at the temperature much larger
than the calculated above
 transition temperature $T_c$ ($\propto 1/l$)
 the response remains quasisuperfluid. More precisely, the response
 is controlled by the rate of phase slip (PS)
events, which are processes by which the superflow dissipates. But
 the relaxation time $\tau$ can be exponentially large, and at time interval much smaller than
  $\tau$ the system behaves as the superfluid one.

The consideration \cite{35, 45} we follow was done with reference to a 2D dislocation
network.

In a uniform 2D system the vortex-antivortex pairs unbind above the BKT transition and
vortices of opposite vorticities can move independently from each other. If a given
vortex crosses the system in a direction perpendicular to the flow, the increase of the
phase of the superfluid order parameter in the flow direction changes on $2\pi$. In a
uniform system a motion of a vortex across the flow is caused by a combined effect of the
Magnus force and a viscous friction force applied to the vortex. In the network, the
vortices are pinned to given plaquettes. Vortexes cannot move freely, but they can "jump"
from one plaquette to another. This process becomes possible due to  emergence of PSC at
the segments.

The PS of the proper sign provides annihilation of the vortex pinned to a given plaquette
and creation of a vortex of the same vorticity in the neighbor plaquette. The process can
be interpreted as a jump. There is no preferable direction of such a jump in a system
without the flow. The situation is changed under the presence of the flow. The rate of
appearance of PSC depends on the superfluid velocity $v_s$ on a segments. The velocity
$v_s$ is the sum of the circular flow connected with the vortex and the uniform flow.
This velocity is different at different segments of the plaquette. As a result, a
preferable direction of the jumps emerges.

In Refs. \cite{35, 45}, the kinetic mechanism of PS was considered. The 1D
Gross-Pitaevskii (GP) equation for the order parameter $\Psi$,
\begin{equation}\label{58}
i\hbar \frac{\partial \Psi}{\partial t}=-\frac{\hbar^2 }{2 m}\frac{\partial^2
\Psi}{\partial x^2}-\mu \Psi+\gamma\lvert\Psi\rvert^2\Psi,
\end{equation}
 has a soliton solution, which describes a density
rarefaction moving through a one-dimensional system.  PS  is connected with the formation
and evolution of  this soliton.

For a condensate that moves with velocity $v_s$, the soliton solution has the form
\begin{eqnarray}\label{59}
\Psi(x,t)=\sqrt{\tilde{n}}
    \Bigg[\sqrt{1-\frac{u^2}{c^2}}\cr
    \times\tanh\left(\sqrt{1-\frac{u^2}{c^2}}\frac{x-(u+v_s)
    t}{\xi}\right) +i\frac{u}{c}\Bigg]e^{\frac{i  m v_s}{\hbar}x},
\end{eqnarray}
where $u$ is the speed of the soliton  relative to the condensate,  $c=\sqrt{\gamma n/M}$
is the velocity of the acoustic mode,  $\xi=\hbar/m c$ is the coherence length,
$\tilde{n}=n/[1-(\xi/l)\sqrt{1-{u^2}/{c^2}}]$, $n$ is the density of the condensate in
the absence of soliton,  and $l$ is the length of the system. One can see from Eq.
(\ref{59}) that the density minimum is the dipper the smaller the speed $u$ is. At $u =
0$ the density goes to zero at the soliton center. The change of the sign of $u$ is
accompanied by an abrupt change on $\pm 2\pi$ of the phase difference at the soliton.

The energy of the soliton is given by the difference of the energy of the system with a
soliton and the energy in its absence.  In the frame of reference in which the fluid is
at rest, the soliton energy is equal to
\begin{eqnarray}\label{60}
    E_{0}=\int_0^{l} d x \left[\frac{\hbar^2}{2
    m}\left(\frac{d \Psi}{d
    x}\right)^2+\frac{\gamma}{2}(\lvert\Psi\rvert^4-n^2)\right]\cr =
    \frac{4}{3}\hbar n c
    \left(1-\frac{u^2}{c^2}\right)^{3/2}.
\end{eqnarray}
The soliton momentum (in the same frame of reference) can be found by integrating the
relation $dp = d E_0/u$ that yields
\begin{equation}\label{61}
    p_{\pm}=-2\hbar n \left(\frac{u}{c}\sqrt{1-\frac{u^2}{c^2}}+
    \arcsin\frac{u}{c}\right)\pm\hbar n \pi.
\end{equation}
 The integration constant is determined from the
condition that $p=0$ at $u=+c$, or $p=0$ at $u=-c$ (at $u=\pm c$ the function
$\Psi(x,t)=\sqrt{n}\exp(i m v_s/\hbar x)$ describes the uniform condensate). These two
conditions give two different integration constants (the second term in Eq. (\ref{61})).
 This ambiguity should be understood as that the momenta
$p_+$ and $p_-$ correspond to solitons nucleated at $u=+c$ and
$u=-c$, respectively. They can be considered as two soliton species
(plus- and minus-solitons). Two soliton species differ from each
other by the shift of the background superfluid velocity $\Delta
v_s$ they induce. Let us explain this point in more detail.

\begin{figure}
\begin{center}
\includegraphics[width=8cm]{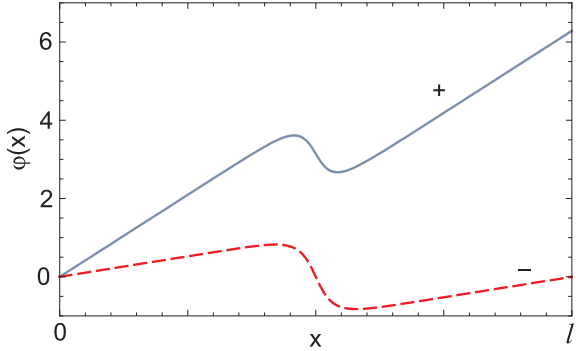}
\end{center}
\caption{The dependence of phase of the condensate wave function on the coordinate in a
system with a plus-soliton (solid curve) and a minus-solitons (dashed curve) at $u=+0.5
c$. Both solitons are centered at $x=l/2$ and were nucleated at the same superfluid
velocity $v_{s0}=2\pi\hbar/m l$.}\label{f4}
\end{figure}

In a multiple connected system the phase satisfies the Onsager-Feynman quantization
condition, and the appearance of a soliton should be accompanied by a change of the net
velocity $\Delta v_s=v_s-v_{s0}$. For a ring with circumference $l$ the shift of the net
velocity is given by the relation
\begin{equation}\label{62}
  \Delta v_{s,\pm} = \frac{\hbar[\pm \pi-2 \arcsin(u/c)]}{m l}
\end{equation}
(the shift is equal to zero at $u=+c$ for plus-soliton, and it is equal to zero at $u=-c$
for minus-soliton). As as example, the dependence of the phase on the coordinate for
plus- and minus-solitons nucleated at the same $v_{s0}$ and reached the same $u$ is shown
in Fig. \ref{f4}. In this figure, the net velocity $v_s$ is proportional to the slope of
the curve $\varphi(x)$ far from the soliton center. One can see that indeed $v_s$ is
different for  plus- and  minus-solitons.

Since $p_+\geq 0$ and $p_-\leq 0$, two exciton species can also be describes as one
specie with the momentum $p$ varied in the range $[-2\pi \hbar n,0)\cup (0 ,  2\pi \hbar
n]$ (at $p=0$ solitons disappear).

 A soliton may change its
momentum in a continuous way due to its interaction with phonons and impurities. The
"diffusion" of solitons in the momentum space is described by the Fokker-Planck equation
for the soliton distribution function $f(p,t)$:
\begin{equation}\label{63}
    \frac{\partial f(p,t)}{\partial t}=-\frac{\partial}{\partial p}\left[A(p)
    f(p,t)-B(p) \frac{\partial f(p,t)}{\partial p}\right].
\end{equation}
The coefficients $A(p)$ and $B(p)$ have the sense of the drift velocity and the diffusion
coefficient in the momentum space.

\begin{figure}
\begin{center}
\includegraphics[width=8cm]{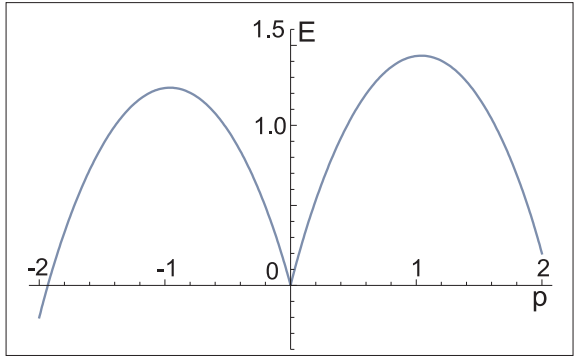}
\end{center}
\caption{The spectrum of solitons $E(p)$ at $v_s=c/(10\pi)$. The energy $E$ is in units
of $\hbar n c$, and the momentum $p$ is in units of $\pi \hbar n$.}\label{f5}
\end{figure}

In the condensate flowing with the velocity $v_s$, the soliton energy is given by the
equation $E(p)=E_0(p)+ v_s p$. The dependence $E(p)$ can be obtained from Eqs. (\ref{60})
and (\ref{61}). The calculated $E(p)$ for $v_s=c/(10\pi)$ is shown in Fig. \ref{f5}. The
function $E(p)$ has two maxima. The "diffusion" over the maximum at $p\approx \pi \hbar
n$ results in $+2\pi$ change of the phase difference
$\Delta\varphi=\varphi(l)-\varphi(0)$, and the "diffusion" over the maximum at $p\approx
- \pi \hbar n$ results in  $-2\pi$ change of $\Delta \varphi$.

The flux of solitons in the momentum space is given by equation
\begin{equation}\label{64}
    s(p)=\frac{1}{2\pi \hbar}\left[A(p)
    f(p,t)-B(p) \frac{\partial f(p,t)}{d p}\right].
\end{equation}
Eq. (\ref{64}) gives the value of flux for a segment of unit length.

 At $p$ close to
zero the distribution of solitons is described by the Bose distribution function. At
$E(p)\gg T$ the Bose distribution can be approximated by $f_0(p)=\exp[-E(p)/T]$. For
$f=f_0$ the flux (\ref{64}) should be equal to zero that yields the relation between
$A(p)$ and $B(p)$:
\begin{equation}\label{65}
   A(p)=-\frac{B(p)}{T} \frac{d E(p)}{d p}=-\frac{B(p)}{T}(u+v_s).
\end{equation}
Considering the stationary flux approximation, one obtains the flux of solitons in the
directions of positive $p$ and negative $p$ (the flux of plus-solitons $s_+$ and the flux
of minus-solitons $s_-$, respectively):
\begin{equation}\label{66}
    s_+=\frac{1}{2\pi \hbar}\frac{1}{\int_0^{2\pi\hbar n}\frac{d p}{B(p)f_0(E(p))}},
\end{equation}
\begin{equation}\label{67}
    s_-=-\frac{1}{2\pi \hbar}\frac{1}{\int_{-2\pi\hbar n}^0\frac{d p}{B(p)f_0(E(p))}}.
\end{equation}
 The negative sign
of the flux $s_-$ indicates that the momentum decreases under "diffusion" over the
corresponding barrier. Approximating the integrals in Eqs. (\ref{66}) and (\ref{67}) by
gauss integrals, one obtains the fluxes as the functions of the superfluid velocity:
\begin{equation}\label{68}
    s_{\pm}(v_s)=\pm\frac{B_{0}}{8\pi \hbar^2 n}\left(\frac{2\hbar n c}{\pi T}\right)^\frac{1}{2}
    \exp\left(-\frac{4 \hbar n c}{3 T}\mp\frac{\pi \hbar n v_{s}}{T}\right),
\end{equation}
where $B_0$ is the diffusion coefficient at the maximum $E(p)$.

The function $A(p)=\langle d p/dt\rangle$ can be evaluated as  the viscous friction force
applied to the soliton. The latter is given by the relation $\langle d
p/dt\rangle\approx-\eta(u+v_s)$, where $\eta$ is the friction coefficient. Taking into
account Eq. (\ref{65}), one finds that $B_0=\eta T$.

For a segment of the length $l$ the  rate of change of the phase difference in the segment,
$\Delta\varphi=\varphi(l)-\varphi(0)$, is given by equation
\begin{equation}\label{69}
    r=\frac{d (\Delta \varphi)}{d t}=2\pi(s_+(v_s)-\lvert s_-(v_s)\rvert)l.
\end{equation}
At $v_s\ll T/(\pi \hbar n)$ this rate is in linear proportion to the velocity $v_s$:
\begin{equation}\label{70}
r=-\frac{\pi \eta l}{2\hbar} v_s\left(\frac{2\hbar n c}{\pi T}\right)^\frac{1}{2}
    \exp\left(-\frac{4 \hbar n c}{3 T}\right).
\end{equation}
For the network, the velocity $v_s$  is the sum of the velocity of the uniform flow and
the velocity of circular flow caused by the vortex.

\begin{figure}
\begin{center}
\includegraphics[width=8cm]{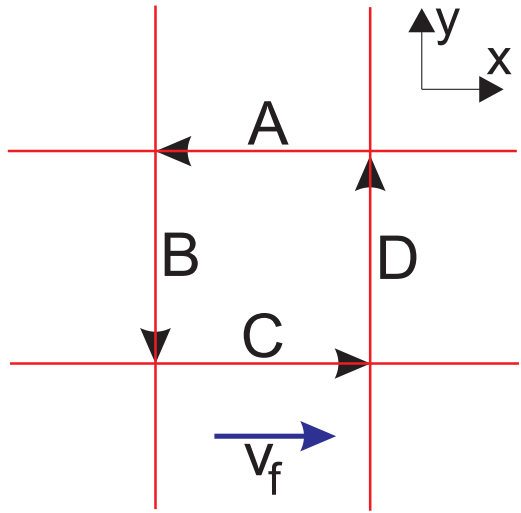}
\end{center}
\caption{Schematic view of the dislocation network plaquette with the vortex in its
center.}\label{f6}
\end{figure}

 We specify a regular quadratic network
with segments oriented along the $x$ and $y$ axes and the uniform flow directed along the
$x$ axis. If the vortex with the positive vorticity is centered in a given plaquette, the
superfluid velocities in the segments that form this plaquette are equal to $v_{s,A}=v_v-
v_f$, $v_{s,B}=v_v$ $v_{s,C}=v_v + v_f$ , and $v_{s,D}=v_v$ (see Fig. \ref{f6}), where
$v_v=\hbar\pi/2 m l$ comes from the vortex velocity field, and $v_f$ is the
velocity of the uniform superflow (we imply that $v_f<v_v$). Since $v_{s,A}<v_{s,C}$, the
 vortexes  with positive vorticity jump predominantly in the $-y$ direction. The average
 velocity of motion of the vortex across the flow is proportional to the difference of
 rates $r$ in the C and A segments:
\begin{equation}\label{71}
    \langle v_{y}\rangle=-\frac{(\lvert r_C \rvert - \lvert r_A \rvert)l}{2\pi}.
\end{equation}
When the vortex crosses the system, the flow velocity $v_f$ changes on $\Delta v_f=-2
\pi\hbar/(m L_x)$.

Vortexes with negative vorticity jump predominantly in the $+y$ direction, move with the
same in module average velocity $\langle v_{y}\rangle$, and their motion also results in
lowering of $v_f$.

The overall rate of change of $v_f$ is given by equation
\begin{equation}\label{72}
    \frac{ d v_f}{d t }=-N_{vort}\frac{2\pi \hbar}{m L_x}\frac{\langle
    v_{y}\rangle}{L_y},
\end{equation}
where $N_{vort}$ is the number of vortexes. Substituting Eqs. (\ref{70}) and (\ref{71})
into Eq. (\ref{72}), we obtain

\begin{equation}\label{73}
    \frac{ d v_s}{d t }=-\frac{v_s}{\tau},
\end{equation}
where
\begin{equation}\label{74}
    \tau = \frac{m}{\eta n_{vort} l^2} \left(\frac{T}{2\pi\hbar n c}\right)^\frac{1}{2}
    \exp\left(\frac{4}{3}
    \frac{\hbar n c}{T}\right)
\end{equation}
and $n_{vort}$ is the vortex density. To estimate $\eta$ one can consider the friction
connected with the interaction of the solitons with phonons. Then, the coefficient $A(p)$
in Eq. (\ref{63}) is evaluated as $A(p)\sim -(mT/\hbar)(u+v_s)$. Thus, $\eta\sim m
T/\hbar$.

The flow velocity decays by the law
\begin{equation}\label{75}
    v_f(t)=v_f(0)e^{-\frac{t}{\tau}}.
\end{equation}
The relaxation time (\ref{74}) is proportional to the exponent $\exp(T_0/T)$, where
\begin{equation}\label{76}
   T_0=\frac{4\hbar n c}{3}.
\end{equation}
At $T\ll T_0$ the time $\tau$ can be extremely large. For instance, at $T=0.03 T_0$ the
relaxation time will be of order or larger than one hour and at $T=0.025 T_0$ it will be
more than one year. The temperature $T_0$ does not depend on the
 length of the segment of the network. It is much larger than the
 temperature of the superfluid transition (\ref{20}). Their ratio is evaluated as
 $T_0/T_c\sim l/\xi\gg 1$. Thus, in the temperature range $T_c\ll T< T_0$ one can expect a
 crossover into a quasisuperfluid state. At
 time intervals much smaller than $\tau$ the behavior of the dislocation network in such a
 state is indistinguishable
 from genuine superfluid.

The picture described in this section is called the Shevchenko state (see, for instance,
\cite{31,33, 46}).

\section{Dislocations with superfluid cores: Numerical and analytical studies}

Two main types of dislocations are  edge dislocations and  screw dislocations. An edge
dislocation is the edge of an extra half-plane inserted into the crystal. A screw
dislocation can be represented as a result of cutting the lattice along the half-plane,
after which the lattice parts on either side of the cut are shifted relative to each
other by one period parallel to the edge of the cut. In a contour enclosed the
dislocation line the elastic displacement vector receives a finite displacement
$\mathbf{b}$ equal in magnitude and direction to one of the lattice vectors. The vector
$\mathbf{b}$ is called the Burgers vector. The Burgers vector of an edge dislocation is
perpendicular to the dislocation line and the Burgers vector of a screw dislocation is
parallel to the dislocation line. Superfluid properties of dislocations may depends on
its type, and on the direction of the Burgers vector and the direction of the dislocation
line relative to the crystallographic axes.

\subsection{Superfluid density and winding number}

The conception of the winding number allows to determine numerically whether a given
dislocation is superfluid or not. In this subsection we follow the paper by Pollock and
Ceperley \cite{47} and obtain the relation between  the winding number and the superfluid
density.

Pollock and Ceperley \cite{47} considered a response of a superfluid system on boundary
motion.  Physically, such a system can be realized  if a superfluid is enclosed between
two cylinders of radii $r$ and $r+d$ rotating with the same angular frequency. For
$d/r\ll 1$ the system becomes equivalent to one enclosed between two planes moving with
velocity $v$ but periodic in one direction.

In the frame at rest with the boundary walls (primed frame) the density matrix operator
of the system  is written as
\begin{equation}\label{77}
    \hat{\rho}'(R,R',\beta)=e^{-\beta H'},
\end{equation}
where $R$ and $R'$ are the variables which denote points in the 3N-dimensional coordinate
space, and
\begin{equation}\label{78}
    H'=\sum_j
\frac{(\mathbf{p}_j-m\mathbf{v})^2}{2 m}+V,
\end{equation}
is the Hamiltonian in primed frame ($V$ is  the potential of particle-particle
interaction). Since the distribution of states is identical in the lab and moving frames,
the density-matrix operator in the lab frame $\hat{\rho}_\mathbf{v}$ is equal to
$\hat{\rho}'$.

The normal component moves with the walls and carries the momentum
\begin{equation}\label{79}
    \langle\mathbf{P}\rangle_\mathbf{v}=\frac{\rho_n}{\rho}mN\mathbf{v}=
    \frac{\mathrm{Tr}[\mathbf{P}\hat{\rho}_v]}{\mathrm{Tr}[\hat{\rho}_v]},
\end{equation}
where $\mathbf{P}$ is the total momentum operator, $\rho_n$ is the normal density, $\rho$
is the total density, and $N$ is the number of particles. The free energy of the system
with moving walls, $F_v$, can be found from the equation
\begin{equation}\label{80}
    e^{-\beta F_v}=\mathrm{Tr}[\hat{\rho}_\mathbf{v}].
\end{equation}

Equation (\ref{79}) can be rewritten as
\begin{equation}\label{81}
\frac{\rho_n}{\rho}mN\mathbf{v}=\frac{1}{\beta}\frac{\partial}{\partial \mathbf{v}}
\ln(\mathrm{Tr}[\hat{\rho}_v])+N m \mathbf{v}=-\frac{\partial F_v}{\partial \mathbf{v}}+N
m \mathbf{v}.
\end{equation}
From Eq. (\ref{81}) one obtains the superfluid density $\rho_s=\rho-\rho_n$:
\begin{equation}\label{82}
    \frac{\rho_s}{\rho}=\frac{\partial\left(\frac{F_v}{N}\right)}
    {\partial\left(\frac{mv^2}{2}\right)}.
\end{equation}
The free-energy change due to the motion of the walls is thus
\begin{equation}\label{83}
\frac{\Delta F_v}{N}=\frac{mv^2}{2}\frac{\rho_s}{\rho}+O(v^4).
\end{equation}

The density matrix $\hat{\rho}_\mathbf{v}$ satisfies the periodic boundary condition
\begin{eqnarray}\label{84}
\hat{\rho}_\mathbf{v}(\mathbf{r}_1,\ldots,\mathbf{r_N};
\mathbf{r}_1,\ldots,\mathbf{r_j}+L,\ldots,\mathbf{r_N},\beta)\cr =
\hat{\rho}_\mathbf{v}(\mathbf{r}_1,\ldots,\mathbf{r_N};
\mathbf{r}_1,\ldots,\mathbf{r_N},\beta),
\end{eqnarray}
where $L$ is the period of the system (circumference of the cylinders).

One can define a modified density matrix
\begin{equation}\label{85}
\hat{\tilde{\rho}}_\mathbf{v}(R,R',\beta)=\exp\left[-i\frac{m}{\hbar}\mathbf{v}
\cdot\sum_j(\mathbf{r}_j-\mathbf{r}'_j)\right]\hat{\rho}_\mathbf{v}(R,R',\beta)
\end{equation}
that satisfies another boundary condition
\begin{eqnarray}\label{86}
\hat{\tilde{\rho}}_\mathbf{v}(\mathbf{r}_1,\ldots,\mathbf{r_N};
\mathbf{r}_1,\ldots,\mathbf{r_j}+L,\ldots,\mathbf{r_N},\beta) \cr
=e^{i\frac{m}{\hbar}\mathbf{v}\cdot\mathbf{L}}\hat{\tilde{\rho}}_\mathbf{v}(\mathbf{r}_1,\ldots,\mathbf{r_N};
\mathbf{r}_1,\ldots,\mathbf{r_N},\beta).
\end{eqnarray}

It follows from  Eqs. (\ref{77}) and (\ref{78}) that the density matrix
$\hat{\rho}_\mathbf{v}$ satisfies the Bloch equation
\begin{eqnarray}\label{87}
    -\frac{\partial \hat{\rho}_\mathbf{v}(R,R',\beta)}{\partial \beta}\cr =\left(\frac{1}{2 m}\sum_j
    (-i\hbar \nabla_j-m \mathbf{v})^2+V\right) \hat{\rho}_\mathbf{v}(R,R',\beta).
\end{eqnarray}
A similar equation for the modified density matrix reads
\begin{equation}\label{88}
    -\frac{\partial {\hat{\tilde{\rho}}}_\mathbf{v}}{\partial \beta}(R,R',\beta)=\left(-\sum_j
    \frac{\hbar^2\nabla_j^2}{2 m}+V\right) \hat{\tilde{\rho}}_v(R,R',\beta).
\end{equation}
This is the equation for the density matrix for a system with stationary walls. Since the
trace over $\hat{\tilde{\rho}}_\mathbf{v}$ is equal to the trace over
$\hat{\rho}_\mathbf{v}$, Eq. (\ref{80}) can be replaces with
\begin{equation}\label{89}
    e^{-\beta F_v}=\mathrm{Tr}[\hat{\tilde{\rho}}_\mathbf{v}].
\end{equation}
Thus, one can use the usual density matrix for a system with stationary walls to
calculate $\hat{\tilde{\rho}}_\mathbf{v}$. The phase factor appeared in the boundary
condition (\ref{86}) should be included as a weight.

\begin{figure}
\begin{center}
\includegraphics[width=8cm]{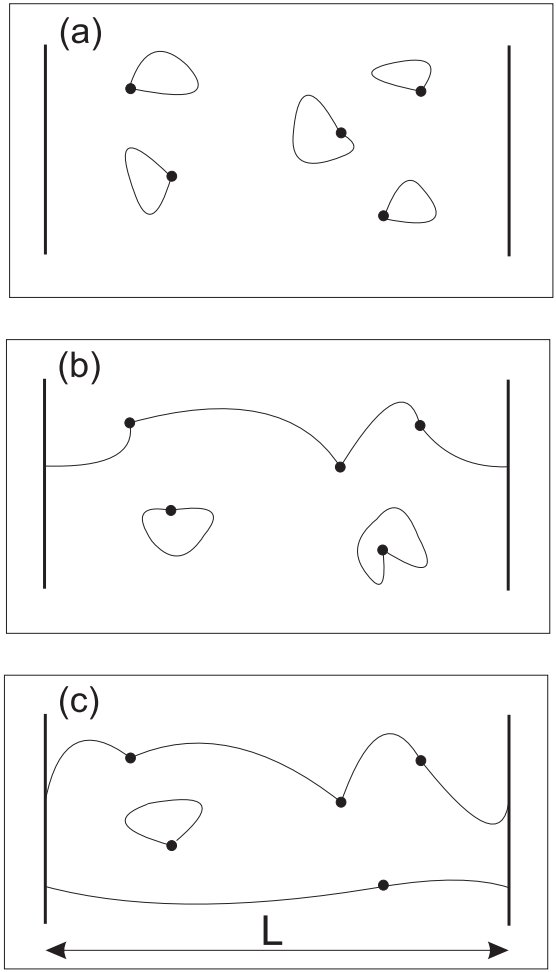}
\end{center}
\caption{Examples of traces of $N=5$ atoms with the winding number (a) $W=0$, (b) $W=1$,
and (c) $W=2$.}\label{f7}
\end{figure}

For periodic boundary conditions the density matrix must include sums over the periodic
images.  For Bose systems an additional sum over permutations is necessary to symmetrize
the density matrix. In a path integral calculation the contribution to the density matrix
from a path ending on a periodic image of its initial point must include the additional
phase factor. This factor expresses through the winding number
\begin{equation}\label{90}
    \mathbf{W}=\frac{1}{L}\sum_{i=1}^N\left(\mathbf{r}_{P_i}-\mathbf{r}_i\right),
\end{equation}
In interpreting Eq. (\ref{90}) one should trace the path of the particle from its origin
at $\mathbf{r}_i$, to its destination at $\mathbf{r}_{P_i}$ and note how many times
periodic boundary conditions have been invoked.  For a periodic cell the winding number
describes the net number of times the paths of the N particles have wound around the cell
(see Fig. \ref{f7}).

The free energy change $\Delta F_v$ is obtained from the winding number distribution as
\begin{equation}\label{91}
   e^{-\beta\Delta F_v}=\frac{\int  \hat{\rho}_\mathbf{v}(R,R,\beta)dR}
   {\int  \hat{\rho}_{\mathbf{v}=0}(R,R,\beta)dR}=\langle e^{i\frac{\hbar}{m}
   \mathbf{v}\cdot\mathbf{W}L}\rangle.
\end{equation}
For  small $v$ Eq. (\ref{91}) yields
\begin{equation}\label{92}
\beta \Delta F_v=\frac{m^2 v^2}{2\hbar^2}\frac{\langle W^2\rangle L^2}{d},
\end{equation}
where $d$ is dimensionality of the system. It follows from the comparison of Eqs.
(\ref{92}) and (\ref{82}) that the superfluid density is proportional to the average
square winding number:
\begin{equation}\label{93}
    \frac{\rho_s}{\rho}=\frac{m}{\hbar^2}\frac{\langle W^2\rangle L^2}{d \beta N}.
\end{equation}

\subsection{Screw and edge superfluid dislocations in solid $^4$He}

\begin{figure}
\begin{center}
\includegraphics[width=8cm]{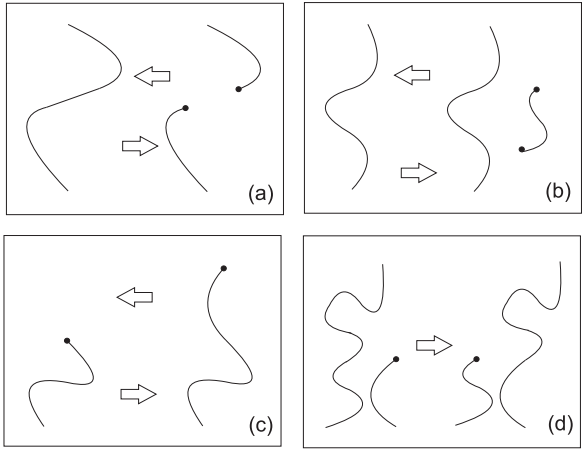}
\end{center}
\caption{Elements of the worm algorithm. (a) opening/closing, (b) inserting/removing, (c)
drawing/erasing, (d) reconnecting. }\label{f8n}
\end{figure}

The average square winding number can be evaluated numerically by path integral Monte
Carlo simulations \cite{47,48}. In the original algorithm \cite{47}, the superfluid
density of liquid $^4$He was obtained by mean of the winding number estimator, which
takes nonzero value if long permutation cycles of identical particles occur in the
system. For closed trajectories, one must insert and remove world lines forming winding
exchange cycles as a whole in a single update. This leads to macroscopically small
acceptance ratios and, thus, severe inefficiency. The algorithm becomes problematic at
large $N$. In the calculations \cite{47}
 the number of atoms was $N=64$. The so-called worm algorithm that allows to perform efficient
 calculations of the superfluid density  with much larger number of atoms was developed
  by Boninsegni, Prokof'ev, and Svistunov
 \cite{49} (see also \cite{46}). In the worm algorithm, open trajectories are allowed.
 The  algorithm includes such elements as erasing  a small part of existing world line (opening) or
 drawing a small piece of world line between the end points to close the loop (Fig. \ref{f8n}(a)),
 inserting a small piece of a new world line or  removing a
small piece of world line between the end points (Fig. \ref{f8n}(b)), drawing or erasing
additional segments of the world line connecting the end points (Fig. \ref{f8n}(c)), and
reconnecting two world lines (Fig. \ref{f8n}(d)). As a result, a finite number of steps
is required to erase any initial trajectory and to draw any other one.

 In
\cite{49} the superfluid transition in liquid $^4$He in 2D was simulated for systems with
up to $N=2500$.

In Ref. \cite{23}, the worm algorithm was used to calculate the superfluid density in a
core of a screw dislocation in solid $^4$He. The simulation sample consisted of an ideal
hcp crystal with a screw dislocation in its center along the c axis. Particles located
far from the dislocation core  had been pinned. These particles were not moved in the
Monte Carlo simulation.  The purpose of "frozen" particles was that of confining the
sample to the inner (cylindrical) part of the cell. The rectangular simulation cell is
twice the size of the physical sample inside the cylinder. The initial number of
(physical) atoms used in simulations was varied from 384 to 1912. The simulations were
performed at the density 0.0287 \AA$^{-3}$, that is, in the close vicinity of the melting
point.

The 1D superfluid density $\rho_s^{(1D)}$ was found from the 1D superfluid stiffness
$\Lambda_s=\hbar^2 \rho_s^{(1D)}/m$. The latter was extracted from the distributions
$P_W(W)$ of the winding along the core, obtained in the simulation,
\begin{equation}\label{94}
P_W(W)\sim \exp\left(-\frac{L W^2}{2\beta \Lambda_s}\right),
\end{equation}
where $L$ is the sample length along the c axis (one can easily check that the
distribution (\ref{94}) yields  $\langle W^2 \rangle$  that satisfies Eq. (\ref{93})).
The obtained $\Lambda_s$ corresponds to $\rho_s^{(1D)}\sim 1$ \AA$^{-1}$ (at $T<1 K$),
which is equivalent to saying that superfluid phase in the dislocation core involves
nearly all atoms in a tube of diameter 6 \AA.

Thus, the screw dislocation in hcp solid $^4$He supports (at $T \to 0$) superfluid
transport of $^4$He atoms along its core.

In Ref. \cite{24}, the results of \textit{ab initio} simulations of edge dislocations in
solid $^4$He at $T=0.5$ K based on the worm algorithm were reported. Since the hcp
structure has two atoms in the unit cell, the edge dislocation involves two extra
halfplanes. This leads to a splitting of the edge dislocation into two partial
dislocations with the fcc stacking fault forming in between. The splitting was so large
that the fault did not fit the simulation cell.  Due to this circumstance simulations of
a single partial attached to the fault were performed. The rectangular simulation cell
contained from 600 to 3400 particles with periodic boundary conditions along the core. In
perpendicular directions, a boundary of pinned $^4$He atoms surrounding a cylinder of
radius $R$ provided the necessary boundary conditions for the simulated sample. Depending
on $R$, the number of actually simulated particles was varied from 270 to 1700. It was
found that the edge dislocation lying in the basal plane and having Burgers vector along
the hcp axis has the superfluid core.

\subsection{Criterion for  dislocation-induced supersolid}

In principle, the coupling of strains with the superfluid order parameter might result in
a transition to the supersolid state even in the absence of dislocations.
 On the phenomenological level such a possibility
 was discussed by  Dorsey, Goldbart, and  Toner \cite{50}. According to Ref. \cite{50},
 the free energy
  which takes into account  the coupling of the
 strain tensor $u_{\alpha\beta}$ with the superfluid order parameter $\psi$
 can be written in
 the form
\begin{eqnarray}\label{95}
    F=\int d^3 r \Big(\frac{1}{2} c_{\alpha\beta}\frac{\partial \psi}{\partial r_\alpha}
\frac{\partial \psi}{\partial r_\beta}+\frac{1}{2}a^{(0)}\lvert\psi\rvert^2\cr +\frac{1}{4!}w
\lvert\psi\rvert^4+ \frac{1}{2}\lambda_{\alpha\beta\gamma\delta}u_{\alpha\beta}u_{\gamma\delta}+
\frac{1}{2}a_{\alpha\beta}^{(1)}u_{\alpha\beta}\lvert\psi\rvert^2\Big).
\end{eqnarray}
The model (\ref{95}) predicted that in a helium crystal the transition to the supersolid
state would be stimulated by internal stresses that make
$a_{\alpha\beta}^{(1)}u_{\alpha\beta}\ne 0$. But two questions were left unanswered: What
type of stress is needed to produce zero-point vacancies in $^4$He and is the strength of
this stress realistic? Pollet \textit{et al} \cite{51} addressed this question using
first-principles simulations. The approach of \cite{51} was based on the worm algorithm
\cite{49}.

It was found in Ref. \cite{21} that vacancy and interstitial activation energies (energy
gaps) in hcp $^4$He are quite large ($\Delta_v>13$ K and $\Delta_i>22$ K,
correspondingly)  in the range of densities $n>n_m$, where $n_m=0.0287$ \AA$^{-3}$ is the
melting density. In Ref. \cite{51},  the dependence of $\Delta_v$ and $\Delta_i$ on the
particle density in the range $n<n_m$ was calculated. It was found that the activation
energies are smaller at lower density. They approach zero at the density $n_c=0.025$
\AA$^{-3}$ which is considerably lower than $n_m$. It means that the metastable hcp
crystal remains insulating at solid densities and even beyond. The hydrostatic strain
required to reach the density $n_c$ is about 13.5 \%  that corresponds to unrealistic
underpressure $\Delta P=-25$ bar.

The influence of anisotropic diagonal traceless strain  $2u_{xx}=2 u_{yy}=-u_{zz}=-u_d$
on the vacancy formation gap $\Delta_v$ was also studied in Ref. \cite{51}. It was found
that the gap closes at $u_d=u_{d}^{(c)}=0.1-0.12$. Such a strain corresponds to hardly
achievable stress $\sigma_{zz}\approx 50$ bar.

 By symmetry
there is no linear coupling between superfluid order parameter and shear strain. The
strain dependence of the gap should be quadratic: $\Delta_v=
\Delta_v^{(0)}[1-(u^2_{zx}+u^2_{zy})/u_s^2]$. Numerical simulations \cite{51} gave the
critical value of shear strain as $u_s=0.15$. The corresponding critical shear stress for
closing the gap in the hcp $^4$He is about 35 bar.

Basing on the results of Ref. \cite{51} one can conclude that the homogeneous
strain-induced supersolid phase is not realized in solid helium.

The results obtained in Ref. \cite{51} do not exclude that supersolid phase forms locally
if nonuniform strain close to structural defects (dislocations) exceeds the critical
value and destabilizes parts of the crystal. This condition can be used as a criterion of
dislocation-induced supersolid. The examples are presented in Ref. \cite{51}.

Screw dislocations oriented along the $z$ direction are characterized by a nonzero value
of $u_{zx}$ which can be estimated by  the modulus of the Burgers vector
$b_z=a\sqrt{8/3}$ divided by $4\pi a/\sqrt{3}$, twice the circumference of the circle
going through the atoms closest to the core ($a$ is the lattice period in the basal
plane). The estimated strain is $\sqrt{u^2_{zx}+u^2_{zy}}\approx 0.22$. It exceeds
considerably the threshold value of $u_c=0.15$ found for shear stress. It is in
accordance with the simulations \cite{23} which show that the superfluid density of the
screw dislocation involves nearly all atoms closest to the nucleus.

Edge dislocations in solid $^4$He  optimize their energy by splitting the core and thus
halving the Burgers vector and the strain.  For the partial edge dislocation with core
along the $y$ direction and Burgers vector along $z$ the half-Burgers vector is
$b/2=a\sqrt{2/3}$, and the estimated strain $u_d\approx (b/2)/(2\pi a)=0.13$ is  larger
than the critical strain $u_{d}^{(c)}$. This dislocation should also be superfluid, as
was confirmed by simulations \cite{24}.

For the split-core edge dislocation along $y$ with Burgers vector along $x$, the
half-Burgers vector is $b/2=a/2$, and the strain $u_d=0.08$ is below the threshold value.
According to the proposed criterion, this dislocation should be insulating.

\section{Experimental observation of mass flow in solid Helium}

As was mentioned in the introduction, the experimental evidence for the flow of atoms
through solid $^4$He was obtained in the University of Massachusetts (UMass) experiment
\cite{26}. In more detail, the results of this experiment were presented in Ref.
\cite{52}.

\begin{figure}
\begin{center}
\includegraphics[width=8cm]{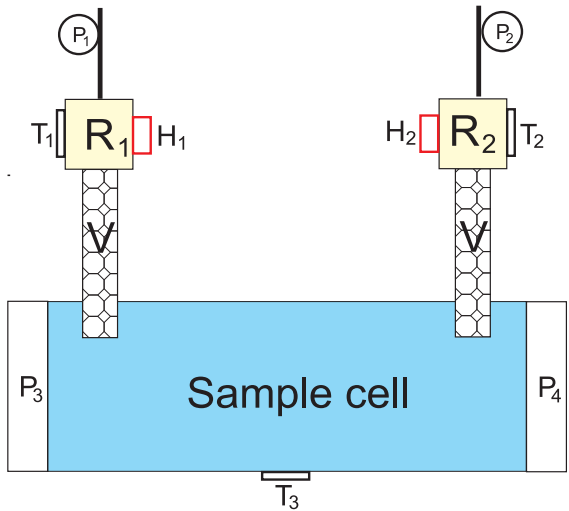}
\end{center}
\caption{Setup of the UMass experiments. The sample cell filled with solid helium
(central chamber) is connected by Vycor glass (V) rods to reservoirs R$_1$ and R$_2$
(outer chambers) filled with superfluid helium. The pressure gauges P$_1$ and P$_2$
measure the pressure in the outer chambers, the pressure gauges P$_3$ and P$_3$ measure
the pressure \textit{in situ} (in the solid sample). The thermometers T$_1$, T$_2$ and
T$_3$ control the temperature of liquid helium in reservoirs and of solid helium in the
cell. The heaters H$_1$ and H$_2$ were used in experiments, where the fountain effect was
utilized.} \label{f8}
\end{figure}

The design of the experiment \cite{26,52} was the following (see Fig. \ref{f8}). Three
chambers were separated from each other by porous Vycor glass. The center chamber
contained the solid hcp helium sample at temperature $T< 1K$, while the outer chambers
and Vycor contained liquid helium (at $T\approx 2$ K in the outer chambers). Below 37 bar
the helium inside the Vycor remained a liquid due to the confined geometry provided by
the pores.  The pressure in outside reservoir filled with liquid helium was larger than
the pressure of solidification of helium in the center chamber ($P_{sol}\approx 25$ bar),
and the contacts of liquid helium with  solid helium were off the melting curve. The
pressure in the outer chambers and on either side of the cell with solid sample was
monitored. A chemical-potential difference between the outer chambers with superfluid
helium was created by injection of helium atoms into one of chambers that resulted in
difference of pressure $\Delta P$ between outside reservoirs. The created pressure
difference decreased with time if the temperature of the solid $^4$He was below
approximately 550 mK and the pressures  was below approximately 26.9 bar. The decrease of
$\Delta P$ indicated that atom flowed through the solid sample. The flow rate
(proportional to the rate of change of the induced pressure difference $F=d(\Delta P)/d
t$) was mostly constant over time, and it was independent
 of the driving pressure difference.
The samples thermally cycled to, or above, 550 mK did not support
flow again when cooled down without first subtracting pressure from
one of the fill lines. A pressure increase in the solid itself was
also observed, which indicated that atoms were being added to the
solid to increase its density at the same time as atoms passed
through the solid.

In the subsequent experiments by the UMass group \cite{53, 54} the fountain effect was
used to create chemical potential differences across samples of solid helium. The design
was basically the same as in Refs. \cite{26,52}, but the flow was induced by the
difference of temperatures of superfluid-filled chambers. This difference resulted in the
appearance of the fountain pressure between the reservoirs and the solid in the sample
cell. The rate $F=d(\Delta P)/d t$ increased smoothly from zero near 600 mK as the
temperature of the solid helium reduced, but at $T\approx 75$ mK it decreased abruptly
and then increased again at lower temperature. The \textit{in situ} pressure gauges
registered a decrease of pressure in the solid by  2.5 mbar in the flow regime, which
indicated that the solid was also a source of atoms.

The same as in Refs. \cite{53, 54} experimental setup was used by the UMass group
\cite{55,56} to measure the dependence of $F$ on the difference of the chemical
potentials $\Delta \mu$ between two reservoirs with liquid helium.  The difference
$\Delta \mu$ was calculated as $\Delta \mu=m\int(d P/ \rho-s d T)$, where $m$ is the
$^4$He mass, $\rho$ is the mass density, and $s$ is the entropy per unit mass. The
obtained dependence $F(\Delta \mu)$ differs from linear one. It was found that a single
function $F=F_0 (\Delta\mu)^b \ln(T/T_0)$, where $b=0.29\pm 0.01$ and $T_0=0.63 \pm 0.01$
K,  fits the data in the range $P=25.6 - 25.8$ bar reasonably well.

In the UMass experiment by Vekhov, Mullin and Hallock \cite{57}, the influence of the
concentration of $^3$He impurities $x_3$ on the temperature  dependence of the flux of
atom through the solid $^4$He was studied. It was established that the temperature,
$T_d$, at which the flux decreases sharply, is an increasing function of $x_3$, and
varied from $T_d\approx 80$ mK at $x_3=0.17$ ppm to $T_d\approx 115$ mK at $x_3=120$ ppm.
At temperatures above $T_d$ the flux has a universal temperature dependence and the flux
terminates in a narrow window near a characteristic temperature $T_h=625$ mK, which is
independent of $x_3$.

\begin{figure}
\begin{center}
\includegraphics[width=8cm]{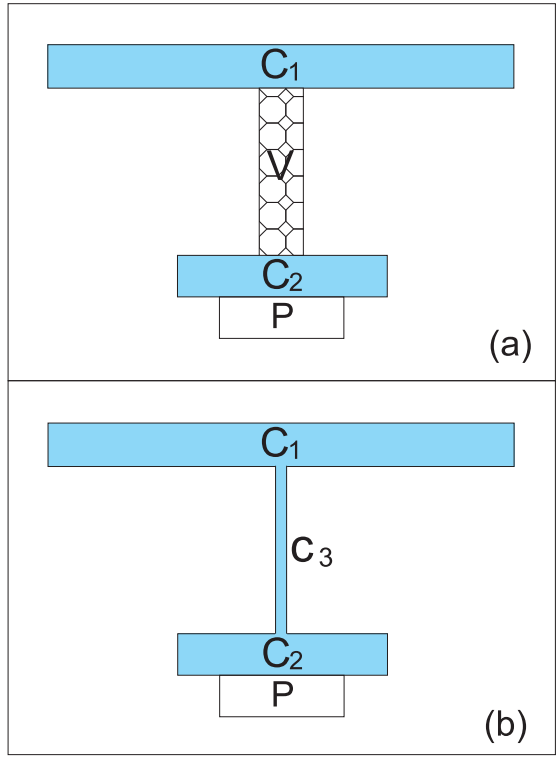}
\end{center}
\caption{Setup of the University of Alberta and ENS experiment (a) and the University of
Alberta experiment (b). C$_1$ is the squeezing chamber filled with solid helium,  C$_2$
is the detecting chamber filled with solid helium, V is the Vicor road filled with liquid
helium, C$_3$ is the central channel filled with solid helium, $P$ is the pressure
gauge.}\label{f9}
\end{figure}

The experimental setup different from the setup of the UMass group was used in the
University of Alberta and ENS experiment \cite{58} (see Fig. \ref{f9}(a)). Two chambers
filled with solid $^4$He were connected by a superfluid Vycor channel. It was observed
that mechanically squeezing the solid in one chamber produced a pressure increase in the
other chamber. For the solid sample at high temperature (1.45 K) thermally activated
vacancy diffusion ensured pressure equilibrium throughout the cell. The thermally
activated pressure response became slower and smaller as the temperature decreased and
disappeared by 700 mK. Flow reappeared below 600 mK. This pressure response was very
similar to the flow seen in the UMass experiments: it increased as the temperature was
reduced, then decreased dramatically at the temperature, $T_d=30 - 140$ mK, depending on
the $^3$He impurity concentration. The observed response was associated with mass
transport through the solid-superfluid-solid junction. The maximum pressure response
below 600 mK was in ten times smaller than the pressure change at high temperature,
indicating that flow does not equilibrate the pressure in the entire squeezing chamber.

In the University of Alberta experiment by Cheng and Beamish \cite{59}, mass flow in
solid helium was observed in a setup in which Vycor was eliminated (see Fig.
\ref{f9}(b)). The experimental cell consisted of two thin disk-shaped chambers connected
by a thin cylindrical flow channel. The chambers and the channel were filled with
polycrystal solid $^4$He. Squeezing of one chamber produced an immediate relatively small
pressure jump, which reversed when the compression was released. This was the result of
elastic deformation of the helium in the flow channel, not of flow through the channel.
Following the elastic jump, the pressure rose more slowly but stopped abruptly after a
few minutes. This was in contrast to the exponential pressure relaxation that is seen for
viscous flow or at temperatures near melting where thermally activated diffusion is
possible. The flow rate was assumed to be proportional to the pressure rise rate. Flow
appeared below 600 mK and increased down to the lowest temperature. The crystals were
grown with $^3$He concentration as low as $x_3=5\times 10^{-12}$. It allowed to avoid the
low temperature flow suppression observed in previous experiments. It was found that the
flow rate continued to increase down to at least 28 mK without saturation. No low
temperature flow drop were found in the sample grown from helium with 120 ppb admixture
of $^3$He.
 A drop was observed
below 100 mK with  20 ppm, but the maximum suppression was only 40\% at 40 mK. A sharper
drop (a maximum suppression of 75\%) was observed below 120 mK with 200 ppm. Only  with
1500 ppm  the flow completely suppressed below 100 mK.
 In the samples with
flow, the temperature dependence was consistent, but the magnitude of the flow varied
between samples and changed when a sample was thermally cycled up to $T\approx 600$ mK.

In the experiment by the Pennsylvania State University (PSU) group \cite{60} the setup
was similar to one of  the UMass group. The main difference was that the solid sample was
of 8 $\mu$m thick. In this sample the dislocation lines were likely pinned at the two
flat surfaces of the solid sample and aligned primarily along the flow path direction
without forming a network. Two different procedures were used to induce flow through the
samples. In the first method, the pressure difference was created by injection of $^4$He
via a capillaries into the left chamber with liquid helium.  In the second method, the
fountain pressure was used. In both methods, a linear decrease in pressure in the left
chamber and the matching linear increase in pressure in the right chamber with liquid He
was observed. It indicated a constant left-to-right mass flow through the solid
independent of the left-right pressure difference. The flow rate decreased with
temperature and decayed nearly exponentially with the pressure of the samples. Any sharp
cutoff in the mass flow rate at low temperature was not observed. The mass-flow rate was
sample dependent and for some samples the enhancement of this rate after thermal
annealing was observed. No evidence of any mass flow extinction or even reduction at low
temperature was found in samples grown with helium gas of $^3$He concentration $x_3$ up
to $1.5 \cdot 10^{-2}$.

\begin{figure}
\begin{center}
\includegraphics[width=8cm]{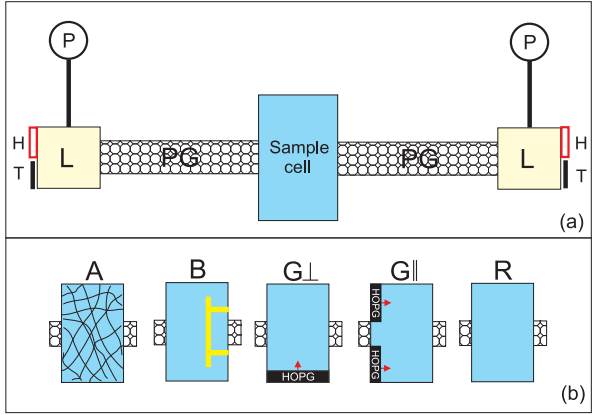}
\end{center}
\caption{(a) Setup of the experiment \cite{61}, and (b) five different sample cells used.
L are the reservoirs filled with liquid helium, PG are the porous glass rods, H are the
heaters, T are the thermometers, P are the pressure gauges, and HOPG is highly oriented
pyrolytic graphite.}\label{f10}
\end{figure}

In the PSU experiment by Shin and Chan \cite{61} five different solid sample cells,
labelled as A, B, G$\perp$, G$\parallel$, and R, were used (see Fig. \ref{f10}). Two
porous glass rods that served as superfluid leads were inserted into the opposite ends of
each of the sample cells. The sell A was completely filled with silica aerogel of 95\%
porosity. The silica strands in the aerogel were randomly interconnected with a mean
separation of 100 nm, orders of magnitude shorter than the typical loop length of a
dislocation network. As a consequence, a dislocation network was not formed. No mass flow
was observed in samples grown in the cell A. No mass flow was also observed in samples in
the  sell B which had a barrier in the form of a thin copper foil suspended in the center
of the cell. Sample cells G$\perp$ and G$\parallel$ were installed with highly oriented
pyrolytic graphite with the $c$ axis aligned respectively perpendicular and parallel to
the flow direction to seed hcp single crystals of $^4$He with the same alignment. The
sample cell R did not contain any inclusions. Mass flow was found in the samples grown in
R, G$\perp$, and G$\parallel$ cells. The measured flow rates at 0.1 K of all samples were
comparable and clustered between 5 and 25 ng/s per mm$^2$. The thicknesses of the samples
were 2.5, 1.9, and 2.4 mm  for the R, G$\perp$ and G$\parallel$ sells, respectively. From
the comparison of the mass flow in that samples with one in 8 $\mu$m sample \cite{60} and
in 2 cm UMass samples (data obtained in Ref. \cite{62}) the authors concluded that the
flow rate decays logarithmically with the thickness of the solid samples.

In the subsequent paper \cite{63} Shin and Chan studied the mass flow rate reduction in
the sample grown from helium gas with different concentration of $^3$He impurities (the
concentration varied in the range from $3\cdot 10^{-7}$ to $10^{-2}$). The sample
thickness was 2.5 mm. The flow rate drop at low temperatures was observed at
$x_3=3.5\cdot 10^{-4}$ and at larger $x_3$. This critical concentration is much larger
than  the concentration in the helium gas used to prepare 2 cm UMass samples \cite{57,
62}. Shin and Chan argued that a very large fraction of the $^3$He atoms is trapped
inside the pores of the Vycor glass and the actual $x_3$ of the solid samples in the
experiment \cite{63} (and also in the UMass experiments \cite{57, 62}) was many orders of
magnitude lower than the $x_3$ of the starting helium gas.

In summary, the experiments of the University of Massachusetts group, the University of
Alberta and ENS groups, and the Pennsylvania State University  group confirmed that
dislocations may provide superflow of atoms through the solid helium as was predicted in
\cite{22,35}.
 The flow was observed at the pressure less than $\approx 30 $ bar (or
smaller, depending on temperature), close to the melting curve. It is in correspondence
with the Shevchenko model of dislocation-mediated superflow \cite{22,35}, according to
which superfluid flow along dislocations is possible due to lowering of local pressure in
the dislocation core. In all mentioned experiments the mass flow was observed below 600
mK and increased with lowering in temperature. The temperature $T_h\approx 0.6$ K can be
understood as the temperature, below which the superfluid density in the core of
dislocation is nonzero (the normal density defined by Eq. (\ref{12}) is less than the
total density). The observed flux  was independent of the pressure difference between
outside reservoirs. It can be explained by that the flux was limited by the critical
velocity of the superflow along the dislocation. The increase of the flux under decrease
in temperature can be connected with the increase in the superfluid density and the
increase in the critical velocity. The absolute value of the flux varied significantly
from sample to sample. Thermal cycling resulted in a decrease and extinction of the mass
flow. These features can be connected with different density of dislocations in different
samples and with lowering of this density under thermal treatment.  The suppression of
the mass flow in a sell filled with porous aerogel is explained by that the network of
dislocations is not formed in such a sample. The lowering and extinction of the mass flow
at large $^3$He concentration at low temperature can be explained by that $^3$He isotopes
are concentrated at the nodes of the dislocation network. For typical length of the
segment of the dislocation of order of 10 $\mu$m the critical temperature  given by Eq.
(\ref{20}) or (\ref{21}) is much smaller than T=0.6 K. Therefore, the mass flow can be
interpreted as a manifestation of the quasisuperfluid state in the dislocation network
(Sec. \ref{s3}).

\section{Superclimb of dislocations and the syringe effect}

A crystal is isochorically compressible if its density $n$ demonstrates a response to
small changes of the chemical potential $\mu$, that is, $\chi=(d n/ d \mu)_V\ne 0$.
Nonzero isochorical compressibility of $^4$He crystal was registered in the UMass
experiments \cite{26, 52}. The effect was observed in the same samples in which the mass
flow was registered. One can say that superflow and isochorically compressibility
accompany each other. Evidence of nonzero $\chi$ was also obtained in the University of
Alberta and ENS experiments \cite{58, 59}, in which superflow of atoms from one solid
sample to the other solid sample was induced by squeezing of one of the samples. The
mechanism of isochorical compressibility of solid helium was proposed by Soyler
\textit{at al} in Ref. \cite{24}. It was argued that the effect is caused by the
superclimb of edge dislocations, controlled by superfluid flow along the dislocation
cores.

The model \cite{24} is based on a coarse-grained description of edge dislocations.
Superfluid edge dislocations  oriented  along the $x$ axis (in the basal plane) with the
Burger vector oriented along the $z$ axis ($C_6$ axis) were considered.  The climb was
described by the core displacement $\delta y(x,t)$. The displacement is perpendicular to
the Burgers vector and should be accompanied with the flux of atoms along the dislocation
toward or away the local displacement. Variation of the linear density in the core is
proportional to core displacement variation: $\delta n_1(x,t)=\xi \delta y(x,t)$, where
$\xi=(a'a)^{-1}$, $a$ is the length of the unit cell along the core, and $a'$ is the
lattice period in the climb direction.
 The linear density variation $\delta n_1$ can be  considered as the conjugate   variable
  to the
phase  $\varphi$ of the superfluid order parameter. Then,  the low-energy effective
action reads
\begin{equation}\label{96}
    S=\int_0^\beta d\tau \int d x \left[-i \hbar\xi y \dot{\varphi}
    +\frac{\hbar^2\rho_s}{2 m}\left(\frac{\partial \varphi}{\partial x}\right)^2-\mu \xi
    y\right]+S_d,
\end{equation}
where $\rho_s$ is the superfluid stiffness, and
\begin{equation}\label{97}
    S_d=\int_0^\beta d\tau \int d x \left[\frac{n_d v_d^2}{2}\left(\frac{\partial y}{\partial
    x}\right)^2-u\cos\left(\frac{2\pi y}{a'}\right)\right]
\end{equation}
is the purely dislocation part taken in the form of the action of a string subjected to
Peierls potential. In Eq. (\ref{97}), $n_d$ is the linear mass density of the string,
$v_d$ is the speed of sound along the string, and $u$ is the strength of Peierls
potential.

Peierls barrier is relevant at low $T$, that leads to a finite gap for the climb motion.
One could expect that the dislocation in its ground state lies in a single Peierls valley
(the limit of smooth dislocation). Then,  Peierls potential can be expanded in powers of
$y$ around given $y_m=m a'$ ($m$ is integer), and the gradient in the action $S_d$ can be
ignored. As a result, the system is described by the standard 1D superfluid action
\begin{equation}\label{98}
    S_1=\int_0^\beta d\tau \int d x \left[-i \hbar\xi y \dot{\varphi}
    +\frac{\hbar^2\rho_s}{2 m}\left(\frac{\partial \varphi}{\partial x}\right)^2-\mu \xi
    y+\frac{g}{2} y^2\right]
\end{equation}
with $g=(2\pi)^2 u/a'^2$. The climb response to $\delta\mu$
  is  $\delta y=\xi \delta\mu/g$, and the density variation is $\delta n_1=\xi^2 \delta\mu/g$. The
  average 3D density  variation can be evaluated as $\delta n= \delta n_1 n_{dis}$,
  where $n_{dis}$ is the density
  of dislocations. One can see that in the limit of smooth dislocations
   the compressibility $\chi$ should be small due to smallness
  of $n_{dis}$  in real samples ($n_{dis}\sim 10^6$ cm$^{-2}$).

At high $T$ Peierls potential is less relevant and can be ignored. In this case, the
dislocations become rough  and the spatial gradient in Eq. (\ref{97}) should be taken
into account. The effective action then reads
\begin{eqnarray}\label{99}
    S_2=\int_0^\beta d\tau \int d x \Bigg[-i \hbar\xi y \dot{\varphi}
    +\frac{\hbar^2\rho_s}{2 m}\left(\frac{\partial \varphi}{\partial x}\right)^2\cr -\mu \xi
    y +\frac{n_d v_d^2}{2}\left(\frac{\partial y}{\partial
    x}\right)^2\Bigg].
\end{eqnarray}
Eq. (\ref{99}) predicts a strong quasistatic climb response: $(\partial_x^2 \delta
y)=-\xi \delta \mu/(n_d v_d^2)$. The response is proportional to the square of the length
$l$ of a segment  of the dislocation network: $ \delta y \sim \xi l^2\delta \mu/(n_d
v_d^2)$. The density variation is given by the expression $\delta n_1 =\xi^2 l^2\delta
\mu/(n_d v_d^2)$. For the network uniform over the whole sample the dislocation density
is $n_{dis}\sim 1/l^2$.  In this case, the  variation of the average 3D density of atoms
$\delta n= \xi^2 \delta \mu/(n_d v_d^2)$ is independent of the density of dislocations,
and the isochorical compressibility $\chi$ is large. Thus, a significant accumulation of
matter in solid $^4$He can be observed. Such an effect was dubbed the giant isochoric
compressibility or the syringe effect.

The model \cite{24} predicted that at low $T$ the superclimb, and, correspondingly, the
syringe effect must be suppressed due to a crossover to a state with smooth dislocations.
But this conclusion was based on the assumption that each dislocation lies in a single
Peierls valley. As was argued in Ref. \cite{64} by Kuklov \textit{et al}, jogs on a
tilted dislocation with superfluid core may form a rough ground state. In this case, the
syringe effect survives down to arbitrarily low temperature in impurity free crystals.

In Ref. \cite{64}, the influence of $^3$He impurities on the  syringe effect was also
discussed. The superflow suppression observed at low temperature \cite{57} can be
explained by blocking of the dislocation cores by the $^3$He impurities, which restrict
the motion of $^4$He atoms along the cores. At the macrolevel, pinning of dislocations by
$^3$He atoms reduces the effective length of the dislocation segments, and  the parameter
$l$ in the expression for the core displacement $\delta y$ is replaced with a much
smaller length $l_p$. This may reduce compressibility $\chi$ and suppress the syringe
effect.

Kuklov \cite{65} proposed to measure the syringe effect in response to the shear stress
to confirm the superclimb scenario. The edge dislocation with the superfluid core and
parallel to it basal edge dislocation with the insulating core interact with each other
through elastic fields. The Burgers vector $\mathbf{b}_c$ of the superfluid edge
dislocation is directed along the C$_6$ symmetry axis. This dislocation can superclimb
along the basal plane (the plane perpendicular to the C$_6$ axis) with the help of extra
matter supplied along its superfluid core. The Burgers vector $\mathbf{b}$ of the
insulating basal dislocation lies in the basal plane. The insulating dislocation can
glide along the basal plane without any need for extra matter.

The stable equilibrium relative positions of the superlimbing and insulating dislocations
can be found as follows. Let one chooses the axis $x$ along $\mathbf{b}$, the axis $y$
along the dislocation lines, and the axis $z$ along $\mathbf{b}_c$. The $x$-component of
the force due to the superclimbing dislocation applied to the insulating dislocation is
equal to $f_x^{(i)} = -b \sigma^{(s)}_{zx}$, where $\sigma^{(s)}_{zx}$ is the stress
tensor due to the superclimbing dislocation. The value this force is equal  to \cite{66}
\begin{equation}\label{100}
    f_x(x,z)=\frac{b b_c G}{2\pi(1-\nu)}\frac{z(x^2-z^2)}{(x^2+z^2)^2},
\end{equation}
where $G$ is the shear modulus, $\nu$ is the Poisson ratio, $z$ and $x$ are the distances
between the dislocations along the C$_6$ axis and in the basal plane, correspondingly.

\begin{figure}
\begin{center}
\includegraphics[width=8cm]{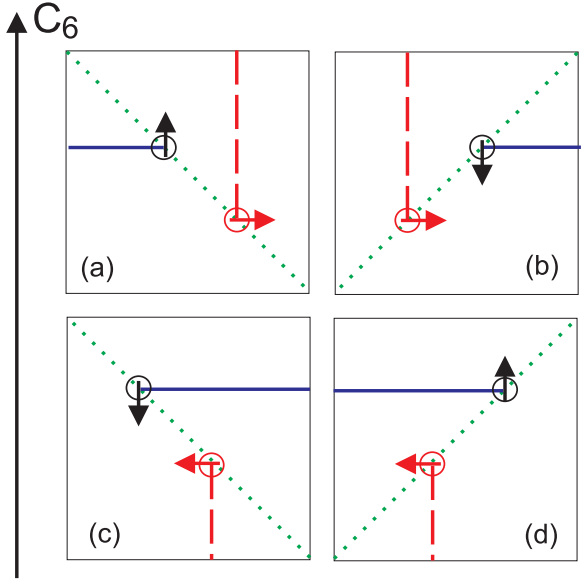}
\end{center}
\caption{Equilibrium positions of the superclimbing (blue) and basal insulating (red)
dislocations relative to the directions of their Burgers vectors (a-d). The additional
halfplane of atoms, which corresponds to the superclimbing (insulating) dislocations is
shown by solid (dash) line. Burgers vectors are shown by arrows. Dislocations cores
(shown by circles) are aligned into the page.}\label{f11}
\end{figure}

The dislocations can move along the $x$ direction only. The distance $z$ is fixed. The
potential energy per unit length $V (x) = - \int_0^x dx' f_x(x',z)$ reads
\begin{equation}\label{101}
    V(x)=\frac{b b_c G}{2\pi(1-\nu)}\frac{z x}{x^2+z^2}.
\end{equation}
The minimum of $V(x)$ is reached at $x = \pm z$ (depending on the signs of $b$ and
$b_c$). Thus, the equilibrium relative positions of the cores of dislocations are located
on straight line which is inclined at 45$^\circ$ with respect to the Burgers vectors
$\mathbf{b}$ and $\mathbf{b}_c$ (see Fig. \ref{f11}).

The binding energy $E_b=b b_c G/[4\pi(1-\nu)]$ is independent of the distance between the
dislocations. An external stress $\sigma^{(ex)}_{zx}$  produces force $f^{(ex)}_x =
b\sigma^{(ex)}_{zx}$ on the insulating dislocation. The potential energy of the pair of
dislocations becomes $V^{(ex)}(x) = V (x) - f^{(ex)}_x x$. Formally,  any external force
breaks the pair. However, a potential barrier which prevents such breaking remains finite
at subcritical  force ($-2E_b/\lvert z\rvert <f^{(ex)}_x<E_b/(2\lvert z \rvert)$). Subcritical force applied to
the insulating dislocation induces superclimb of the dislocation with the superfluid
core. Thus, plastic deformation should induce the syringe effect. The inverse effect,
inducing of plastic deformation by injection of atoms in the sample also will take place.

\begin{figure}
\begin{center}
\includegraphics[width=6cm]{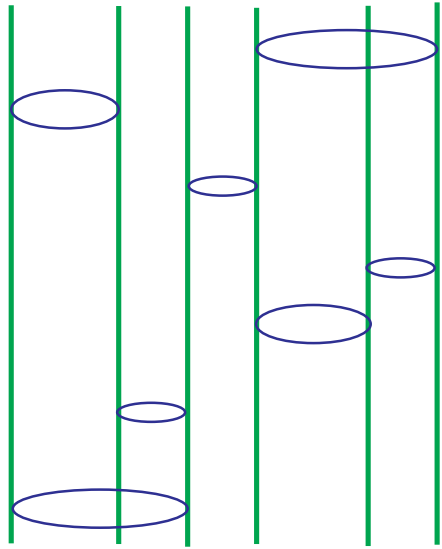}
\end{center}
\caption{The schematic view of possible superfluid network of dislocations. Vertical
lines represent superfluid screw dislocations linked to the superfluid edge dislocations
forming prismatic loops.}\label{f12}
\end{figure}

In Ref. \cite{33}  Kuklov, Prokof'ev, and Svistunov  proposed a model of a 3D network
consisting of a forest of screw dislocations pinned by prismatic loops made of superfluid
edge dislocations (Fig. \ref{f12}). The problem of stability of such a network was
addresses by the same authors in Ref. \cite{67}. A prismatic loop is  a tight cluster of
vacancies or interstitials in the basal plane surrounded  by a rim of partial edge
dislocations with the (half) Burgers vector directed along the C$_6$ axis. This rim is
superfluid, and the loop can move ballistically by dissipationless matter transfer from
one end of the loop to the other. At large distances, prismatic loops interact by sign
varying long-range forces similar to dipole forces. In particular, two loops belonging to
the same basal plane and characterized by the same Burgers vector repel each other.
However, at a distance comparable or smaller than loop length $R$, the repulsion is
changed to attraction, and loops merges together to form a larger loop with the total
length reduced from $2R$ to $\sqrt{2}R$. The lowest energy state of a dilute system of
$N$ loops corresponds to one macroscopic loop with size $\propto \sqrt{N}$. The best
metastable configuration is a dipole solid with inter-loop separation $D > R$. The
leading destabilizing mechanism is particle transfer between the loops. Even in the
absence of particle transfer within the percolating network, tunneling of individual
atoms between the loops makes this configuration unstable.

The described scenario does not take into account that segments of superclimbing
dislocations can be trapped by  insulating dislocations. It was suggested in Ref.
\cite{67} that a stable network of insulating dislocations can stabilize a network of
superclimbing dislocations. There are two options: (i) long superclimbing dislocations
separated from each other and trapped by the network of insulating dislocations; (ii)
mesoscopic prismatic loops stabilized by the network of insulating dislocations. In the
latter case, the percolation of the flow is established due to Josephson effect between
neighboring loops.

In the case (i) the  superflow is determined essentially by
properties of long superclimbing  dislocations, while  the syringe
effect is determined by a typical size of bound segments of
superclimbing and insulating dislocations.

In the case (ii) the superflow will be dominated by the Josephson junctions between the
trapped prismatic loops. In this case, a macroscopic number of such loops should be
formed close enough to each other to provide the superflow. But it is highly unlikely to
occur during random process of solid growth. Nevertheless,  if the conducting pathways do
not formed under the crystal growth,  they can be formed in response to  the injection of
$^4$He atoms into the solid. These atoms form prismatic loops and, as was shown by Kuklov
\cite{68}, such loops become unstable against inflation for large enough chemical
potential bias.
 Injected loops may grow macroscopically large and establish the superflow
pathways through the whole sample. Detecting the threshold for inducing the superflow and
the syringe effects will be a strong indication for such a mechanism.

In short samples,  the flow can be supported by straight screw dislocations. As was
argued in Ref. \cite{68}, screw dislocations can develop helical instability which should
also lead to the syringe effect. In this case, there also should be a threshold for the
chemical potential bias inducing the syringe response.

\section{Conclusion}

The Andreev-Lifshitz theory of the vacancy supersolid and the Shevchenko theory of the
dislocation supersolid were phenomenological. They were based on  reasonable assumptions,
but only experiments could prove or refute the validity of these assumptions.

The supersolidity could be confirmed  by measurements of the moment of inertia of a
rotating crystal. A drop of the moment of inertia below a certain temperature would
indicate the appearance of a superfluid fraction in the system. Similar experiment with
superfluid helium is known as the Andronikashvili experiment. In the Kim and Chan
experiment and in a number of experiments by other groups, the measurements of the period
of a torsional oscillator were done with the goal to find such a drop. Although a sharp
decrease in the oscillation period with decreasing temperature was indeed observed, as
further investigation showed, the decrease in the oscillation period was connected with a
change in the torsion stiffness of the rod, and in the experiment where the torsion
stiffness was controlled, no such decrease was observed.

The negative result of experiments with the torsional oscillator does not rule out the
possibility of the dislocation supersolid. For helium crystal with superfluid
dislocations the expected change in the moment of inertia is too small to register in
torsional oscillator experiments. In contract, experiments on mass transport may register
the dislocation supersolid. This is a system in which a helium crystal is connected to
two reservoirs filled with superfluid helium. The mass transport through a crystal may
emerge at nonzero gradient of the chemical potential of atoms. The
 gradient can be created by applying a pressure difference between the reservoirs.
The idea of such an experiment was put forward in the papers by Shevchenko \cite{22,35}.
The important improvement by Ray and Hallock \cite{26} was that liquid helium at the
boundary with solid helium should be in a microporous environment. The reason is that, in
a microporous environment, helium solidifies at a higher pressure than bulk helium
solidifies and the gradient of the chemical potential along the flow path can be created.
The experiment was proposed by Svistunov and Prokof'ev (see, the review \cite{29}).

 Ray
and Hallock used Vycor rods to connect reservoirs with liquid helium  to a cell with
solid helium. It was found that the addition of helium atoms to one of the reservoirs
causes an increase in pressure in the other reservoir, indicating the appearance of flow
of helium atoms through the solid sample. The effect was associated with the presence of
 dislocations  with superfluid cores in the solid.
  It was found that the mass flow is accompanied by an increase in pressure in the
solid sample, which indicates the accumulation of some of  added atoms in this sample,
i.e., an increase in the density of the sample. The increase in the density of the solid
sample was associated with climb of dislocations in the direction perpendicular to the
Burgers vector. Such climb requires additional atoms to enter the sample. Climb of
dislocations occurs rapidly because of superflow of vacancies or interstitial atoms along
the dislocations. The increase in the density of a solid sample in the presence of
superfluid dislocations was called the syringe effect \cite{24}.

The experiment by Vekhov, Mullin and Hallock \cite{57} had demonstrated that $^3$He
impurities suppress the superflow of atom through solid helium at low temperature. Shin
and Chan \cite{63} established that the critical concentration of $^3$He in the initial
gas, at which the mass flow stops, increases with decreasing the sample thickness, and in
very thin samples, in which dislocations adjoin the sample without intersections, no flow
suppression was detected at all up to very large concentrations of $^3$He. This behavior
allows one to relate the suppression of the flow to the accumulation of $^3$He atoms on
the intersections of dislocations. It was found by Shin and Chan \cite{61} that solid
helium grown in a porous medium, which prevents the formation of dislocation paths
through the sample, does not demonstrate any superflow effect. It was also shown in Ref.
\cite{61} that the value of the flux does not depend on its direction relative to the
crystallographic axes, i.e., both dislocations lying in the basal plane and dislocations
directed along the hexagonal symmetry axis provide the mass flow.

 The experimental results obtained give sufficient support for the theory
of dislocation supersolid by Shevchenko as applied to $^4$He crystals.

In conclusion, we would like to cite R. Hallock. In his recent review \cite{29} he
concluded that "...  while solid $^4$He shows little or no evidence for the presence of
supersolidity of the type proposed nearly 50 years ago (\cite{1}), it provides us with
strong evidence for an entirely different type of superflow behavior reminiscent of what
was initially proposed by Shevchenko."
\section*{Compliance with Ethical Standards}
 \subsection*{Confict of interest} The author declares that he has no confict of interest.

\end{document}